\documentclass[12pt,reqno]{amsart}
\usepackage{amsmath,amsthm,amsfonts}

\def\LL{\leavevmode\setbox0=\hbox{L}\hbox to\wd0{\hss\char'40L}}
\def\al{\alpha}

\def\ep{\varepsilon}

\def\la{\lambda}
\def\rh{\rho}

\def\Ga{\Gamma}

\def\P{{\Bbb P}}

\def\today{\ifcase\month\or
 January\or February\or March\or April\or May\or June\or
 July\or August\or September\or October\or November\or December\fi
 \space\number\day, \number\year}
\def\({\left(}
\def\){\right)}
\def\[{\left[}
\def\]{\right]}

\def\3{\ss}

\catcode`\@=11

\thinlines
\newskip\Einheit \Einheit=0.6cm
\newcount\xcoord \newcount\ycoord
\newdimen\xdim \newdimen\ydim \newdimen\PfadD@cke \newdimen\Pfadd@cke
\def\PfadDicke#1{\PfadD@cke#1 \divide\PfadD@cke by2 \Pfadd@cke\PfadD@cke \multiply\PfadD@cke by2}
\long\def\LOOP#1\REPEAT{\def\BODY{#1}\ITERATE}
\def\ITERATE{\BODY \let\next\ITERATE \else\let\next\relax\fi \next}
\let\REPEAT=\fi
\def\Punkt{\hbox{\raise-2pt\hbox to0pt{\hss\scriptsize$\bullet$\hss}}}
\def\DuennPunkt(#1,#2){\unskip
  \raise#2 \Einheit\hbox to0pt{\hskip#1 \Einheit
          \raise-2.5pt\hbox to0pt{\hss\normalsize$\bullet$\hss}\hss}}
\def\NormalPunkt(#1,#2){\unskip
  \raise#2 \Einheit\hbox to0pt{\hskip#1 \Einheit
          \raise-3pt\hbox to0pt{\hss\large$\bullet$\hss}\hss}}
\def\DickPunkt(#1,#2){\unskip
  \raise#2 \Einheit\hbox to0pt{\hskip#1 \Einheit
          \raise-4pt\hbox to0pt{\hss\Large$\bullet$\hss}\hss}}
\def\Kreis(#1,#2){\unskip
  \raise#2 \Einheit\hbox to0pt{\hskip#1 \Einheit
          \raise-4pt\hbox to0pt{\hss\Large$\circ$\hss}\hss}}
\def\Diagonale(#1,#2)#3{\unskip\leavevmode
  \xcoord#1\relax \ycoord#2\relax
      \raise\ycoord \Einheit\hbox to0pt{\hskip\xcoord \Einheit
         \unitlength\Einheit
         \line(1,1){#3}\hss}}
\def\AntiDiagonale(#1,#2)#3{\unskip\leavevmode
  \xcoord#1\relax \ycoord#2\relax \advance\xcoord by -0.05\relax
      \raise\ycoord \Einheit\hbox to0pt{\hskip\xcoord \Einheit
         \unitlength\Einheit
         \line(1,-1){#3}\hss}}
\def\Pfad(#1,#2),#3\endPfad{\unskip\leavevmode
  \xcoord#1 \ycoord#2 \thicklines\ZeichnePfad#3\endPfad\thinlines}
\def\ZeichnePfad#1{\ifx#1\endPfad\let\next\relax
  \else\let\next\ZeichnePfad
    \ifnum#1=1
      \raise\ycoord \Einheit\hbox to0pt{\hskip\xcoord \Einheit
         \vrule height\Pfadd@cke width1 \Einheit depth\Pfadd@cke\hss}%
      \advance\xcoord by 1
    \else\ifnum#1=2
      \raise\ycoord \Einheit\hbox to0pt{\hskip\xcoord \Einheit
        \hbox{\hskip-1pt\vrule height1 \Einheit width\PfadD@cke depth0pt}\hss}%
      \advance\ycoord by 1
    \else\ifnum#1=3
      \raise\ycoord \Einheit\hbox to0pt{\hskip\xcoord \Einheit
         \unitlength\Einheit
         \line(1,1){1}\hss}
      \advance\xcoord by 1
      \advance\ycoord by 1
    \else\ifnum#1=4
      \raise\ycoord \Einheit\hbox to0pt{\hskip\xcoord \Einheit
         \unitlength\Einheit
         \line(1,-1){1}\hss}
      \advance\xcoord by 1
      \advance\ycoord by -1
    \fi\fi\fi\fi
  \fi\next}
\def\hSSchritt{\leavevmode\raise-.4pt\hbox to0pt{\hss.\hss}\hskip.2\Einheit
  \raise-.4pt\hbox to0pt{\hss.\hss}\hskip.2\Einheit
  \raise-.4pt\hbox to0pt{\hss.\hss}\hskip.2\Einheit
  \raise-.4pt\hbox to0pt{\hss.\hss}\hskip.2\Einheit
  \raise-.4pt\hbox to0pt{\hss.\hss}\hskip.2\Einheit}
\def\vSSchritt{\vbox{\baselineskip.2\Einheit\lineskiplimit0pt
\hbox{.}\hbox{.}\hbox{.}\hbox{.}\hbox{.}}}
\def\DSSchritt{\leavevmode\raise-.4pt\hbox to0pt{%
  \hbox to0pt{\hss.\hss}\hskip.2\Einheit
  \raise.2\Einheit\hbox to0pt{\hss.\hss}\hskip.2\Einheit
  \raise.4\Einheit\hbox to0pt{\hss.\hss}\hskip.2\Einheit
  \raise.6\Einheit\hbox to0pt{\hss.\hss}\hskip.2\Einheit
  \raise.8\Einheit\hbox to0pt{\hss.\hss}\hss}}
\def\dSSchritt{\leavevmode\raise-.4pt\hbox to0pt{%
  \hbox to0pt{\hss.\hss}\hskip.2\Einheit
  \raise-.2\Einheit\hbox to0pt{\hss.\hss}\hskip.2\Einheit
  \raise-.4\Einheit\hbox to0pt{\hss.\hss}\hskip.2\Einheit
  \raise-.6\Einheit\hbox to0pt{\hss.\hss}\hskip.2\Einheit
  \raise-.8\Einheit\hbox to0pt{\hss.\hss}\hss}}
\def\SPfad(#1,#2),#3\endSPfad{\unskip\leavevmode
  \xcoord#1 \ycoord#2 \ZeichneSPfad#3\endSPfad}
\def\ZeichneSPfad#1{\ifx#1\endSPfad\let\next\relax
  \else\let\next\ZeichneSPfad
    \ifnum#1=1
      \raise\ycoord \Einheit\hbox to0pt{\hskip\xcoord \Einheit
         \hSSchritt\hss}%
      \advance\xcoord by 1
    \else\ifnum#1=2
      \raise\ycoord \Einheit\hbox to0pt{\hskip\xcoord \Einheit
        \hbox{\hskip-2pt \vSSchritt}\hss}%
      \advance\ycoord by 1
    \else\ifnum#1=3
      \raise\ycoord \Einheit\hbox to0pt{\hskip\xcoord \Einheit
         \DSSchritt\hss}
      \advance\xcoord by 1
      \advance\ycoord by 1
    \else\ifnum#1=4
      \raise\ycoord \Einheit\hbox to0pt{\hskip\xcoord \Einheit
         \dSSchritt\hss}
      \advance\xcoord by 1
      \advance\ycoord by -1
    \fi\fi\fi\fi
  \fi\next}
\def\Koordinatenachsen(#1,#2){\unskip
 \hbox to0pt{\hskip-.5pt\vrule height#2 \Einheit width.5pt depth1 \Einheit}%
 \hbox to0pt{\hskip-1 \Einheit \xcoord#1 \advance\xcoord by1
    \vrule height0.25pt width\xcoord \Einheit depth0.25pt\hss}}
\def\Koordinatenachsen(#1,#2)(#3,#4){\unskip
 \hbox to0pt{\hskip-.5pt \ycoord-#4 \advance\ycoord by1
    \vrule height#2 \Einheit width.5pt depth\ycoord \Einheit}%
 \hbox to0pt{\hskip-1 \Einheit \hskip#3\Einheit 
    \xcoord#1 \advance\xcoord by1 \advance\xcoord by-#3 
    \vrule height0.25pt width\xcoord \Einheit depth0.25pt\hss}}
\def\Gitter(#1,#2){\unskip \xcoord0 \ycoord0 \leavevmode
  \LOOP\ifnum\ycoord<#2
    \loop\ifnum\xcoord<#1
      \raise\ycoord \Einheit\hbox to0pt{\hskip\xcoord \Einheit\Punkt\hss}%
      \advance\xcoord by1
    \repeat
    \xcoord0
    \advance\ycoord by1
  \REPEAT}
\def\Gitter(#1,#2)(#3,#4){\unskip \xcoord#3 \ycoord#4 \leavevmode
  \LOOP\ifnum\ycoord<#2
    \loop\ifnum\xcoord<#1
      \raise\ycoord \Einheit\hbox to0pt{\hskip\xcoord \Einheit\Punkt\hss}%
      \advance\xcoord by1
    \repeat
    \xcoord#3
    \advance\ycoord by1
  \REPEAT}
\def\Label#1#2(#3,#4){\unskip \xdim#3 \Einheit \ydim#4 \Einheit
  \def\lo{\advance\xdim by-.5 \Einheit \advance\ydim by.5 \Einheit}%
  \def\llo{\advance\xdim by-.25cm \advance\ydim by.5 \Einheit}%
  \def\loo{\advance\xdim by-.5 \Einheit \advance\ydim by.25cm}%
  \def\o{\advance\ydim by.25cm}%
  \def\ro{\advance\xdim by.5 \Einheit \advance\ydim by.5 \Einheit}%
  \def\rro{\advance\xdim by.25cm \advance\ydim by.5 \Einheit}%
  \def\roo{\advance\xdim by.5 \Einheit \advance\ydim by.25cm}%
  \def\l{\advance\xdim by-.30cm}%
  \def\r{\advance\xdim by.30cm}%
  \def\lu{\advance\xdim by-.5 \Einheit \advance\ydim by-.6 \Einheit}%
  \def\llu{\advance\xdim by-.25cm \advance\ydim by-.6 \Einheit}%
  \def\luu{\advance\xdim by-.5 \Einheit \advance\ydim by-.30cm}%
  \def\u{\advance\ydim by-.30cm}%
  \def\ru{\advance\xdim by.5 \Einheit \advance\ydim by-.6 \Einheit}%
  \def\rru{\advance\xdim by.25cm \advance\ydim by-.6 \Einheit}%
  \def\ruu{\advance\xdim by.5 \Einheit \advance\ydim by-.30cm}%
  #1\raise\ydim\hbox to0pt{\hskip\xdim
     \vbox to0pt{\vss\hbox to0pt{\hss$#2$\hss}\vss}\hss}%
}
\catcode`\@=12

\font\scalefont = cmti8
 
\thicklines

\newcount\xmin                % coordinates of extremal points of the grid
\newcount\xmax
\newcount\ymin
\newcount\ymax

\newcount\gridwidth
\newcount\gridheight

\newcount\nofxpoints          % number of points in horizontal direction
\newcount\nofypoints          % number of points in vertical direction

\newcount\dcnta               % dummy-counter A
\newcount\dcntb               % dummy-counter B

\def\begingrid#1#2#3#4{			% arguments: xmin, ymin, xmax, ymax
	\global\xmin = #1
	\global\ymin = #2
	\global\xmax = #3
	\global\ymax = #4
	\ifnum\xmin > \xmax\errmessage{PATHS: \xmin > \xmax|}\fi
	\ifnum\ymin > \ymax\errmessage{PATHS: \ymin > \ymax|}\fi
	\gridwidth = \xmax
	\gridheight = \ymax
	\advance\gridwidth by -\xmin
	\advance\gridheight by -\ymin
	\nofxpoints = \gridwidth
	\advance\nofxpoints by 1
	\nofypoints = \gridheight
	\advance\nofypoints by 1
	\advance\gridwidth by 4
	\advance\gridheight by 4
	\advance\xmin by -2
	\advance\ymin by -2
	\begin{picture}(\gridwidth, \gridheight)(\xmin, \ymin)
	\advance\gridwidth by -4
	\advance\gridheight by -4
	\advance\xmin by 2
	\advance\ymin by 2
	\global\dcnta = \ymin
	\loop\ifnum\dcnta<\ymax
		\makegridline\dcnta	% do inner loop in a separate macro
		\global\advance\dcnta by 1
	\repeat
	\makegridline\ymax
}
 
\def\makegridline#1{
   \begingroup
   \global\dcntb = \xmin
   \loop\ifnum\dcntb<\xmax
      \put(\dcntb, #1){\circle*{0.1}}
      \global\advance\dcntb by 1
   \repeat
   \put(\xmax, #1){\circle*{0.1}}
   \endgroup
}
 
\def\makexaxis{\thinlines
        \dcnta = \gridwidth
        \advance\dcnta by 1
        \put(\xmin, 0){\vector(1,0){\dcnta}}
        \thicklines
}
\def\makeyaxis{\thinlines
        \dcntb = \gridheight
        \advance\dcntb by 1
        \put(0, \ymin){\vector(0,1){\dcntb}}
        \thicklines
}
\def\makeaxes{\thinlines
        \dcnta = \gridwidth
        \advance\dcnta by 1
        \put(\xmin, 0){\vector(1,0){\dcnta}}
        \dcntb = \gridheight
        \advance\dcntb by 1
        \put(0, \ymin){\vector(0,1){\dcntb}}
        \thicklines
}
 
\def\makexscale{
   \dcnta = \xmin
	\ifnum\dcnta<-9\dcnta=-9\fi
   \loop\ifnum\dcnta<0
		\put(\dcnta,0){\line(0,-1){0.2}}
      \put(\dcnta,-0.5){\makebox(0,0){
		{\scalefont\number\dcnta}}}
      \advance\dcnta by 1
  \repeat
  \dcnta = 1
  \dcntb=\xmax
  \ifnum\dcntb>9\dcntb=9\fi
  \loop\ifnum\dcnta<\dcntb
      \put(\dcnta,0){\line(0,-1){0.2}}
      \put(\dcnta, -0.5){\makebox(0,0){
		{\scalefont\number\dcnta}}}
      \advance\dcnta by 1
  \repeat
  \put(\dcntb,0){\line(0,-1){0.2}}
  \put(\dcntb, -0.4){\makebox(0,0){{\scalefont\number\dcntb}}}
}
\def\makeyscale{
   \dcnta = \ymin
   \loop\ifnum\dcnta<0
      \put(0,\dcnta){\line(-1,0){0.2}}
      \put(-0.4, \dcnta){\makebox(0,0){
		{\scalefont\number\dcnta}}}
      \advance\dcnta by 1
   \repeat
   \dcnta = 1
   \loop\ifnum\dcnta<\ymax
      \put(0,\dcnta){\line(-1,0){0.2}}
      \put(-0.4, \dcnta){\makebox(0,0){
		{\scalefont\number\dcnta}}}
      \advance\dcnta by 1
   \repeat
   \put(-0.4, \ymax){\makebox(0,0){{\scalefont\number\ymax}}}
}

\newsavebox{\skewdowndottedbox}
\newsavebox{\skewupdottedbox}
\savebox{\skewdowndottedbox}(1,1)[lb]{
	\put(0.2,-0.2){\circle*{0.01}}
   \put(0.4,-0.4){\circle*{0.01}}
 	\put(0.6,-0.6){\circle*{0.01}}
	\put(0.8,-0.8){\circle*{0.01}}
}
\savebox{\skewupdottedbox}(1,1)[lb]{
	\put(0.2,0.2){\circle*{0.01}}
   \put(0.4,0.4){\circle*{0.01}}
 	\put(0.6,0.6){\circle*{0.01}}
	\put(0.8,0.8){\circle*{0.01}}
}

\newsavebox{\hplainbox}
\savebox{\hplainbox}{\put(0,0){\line(1,0){1}}}
\newsavebox{\hdottedbox}
\savebox{\hdottedbox}(1,0)[bl]{
	\put(0.2,0){\circle*{0.01}}
	\put(0.4,0){\circle*{0.01}}
	\put(0.6,0){\circle*{0.01}}
	\put(0.8,0){\circle*{0.01}}
}
\newsavebox{\vplainbox}
\savebox{\vplainbox}{\put(0,0){\line(0,1){1}}}
\newsavebox{\vdottedbox}
\savebox{\vdottedbox}(0,1)[bl]{
	\put(0,0.2){\circle*{0.01}}
	\put(0,0.4){\circle*{0.01}}
	\put(0,0.6){\circle*{0.01}}
	\put(0,0.8){\circle*{0.01}}
}
\newsavebox{\vplaindownbox}
\savebox{\vplaindownbox}{\put(0,0){\line(0,-1){1}}}
\newsavebox{\vdotteddownbox}
\savebox{\vdotteddownbox}(0,1)[bl]{
	\put(0,-0.2){\circle*{0.01}}
	\put(0,-0.4){\circle*{0.01}}
	\put(0,-0.6){\circle*{0.01}}
	\put(0,-0.8){\circle*{0.01}}
}
\newsavebox{\skewupbox}
\savebox{\skewupbox}{\put(0,0){\line(1,1){1}}}
\newsavebox{\skewdownbox}
\savebox{\skewdownbox}{\put(0,0){\line(1,-1){1}}}
\newsavebox{\hstep}		% horizontal step
\newsavebox{\vstep}		% vertical step
\newsavebox{\sstep}		% skew step
\newcount\updownincrement

\def\dosteplist{\afterassignment\handlenextstep\let\next=}
\def\handlenextstep{
	\ifx\next\endList
   	\let\next=\relax
	\else
		\ifx\next-
			\put(\dcnta,\dcntb){\usebox{\hstep}}
			\advance\dcnta by 1
		\else
			\ifx\next|
				\put(\dcnta,\dcntb){\usebox{\vstep}}
				\advance\dcntb by\updownincrement
			\else
				\ifx\next/
					\put(\dcnta,\dcntb){\usebox{\sstep}}
					\advance\dcnta by 1
					\advance\dcntb by\updownincrement
				\else
					\errmessage{PATHS: Wrong symbol path argument.}
				\fi
			\fi
		\fi
		\let\next=\dosteplist
	\fi
	\next
}
 
\def\endgrid{\end{picture}}

\newcount\rowcount
\newcount\columncount

\def\begintableau#1#2{									% rows, columns
\global\rowcount=#1
\begin{picture}(#2,#1)(0,0)
\linethickness{0.1pt}
}

\def\endtableau{\end{picture}}

\def\row#1{
	\global\advance\rowcount by -1
	\global\columncount=0
	\dorowlist#1\endList
}

\def\dorowlist{\afterassignment\handlenextentry\let\next=}

\def\handlenextentry{
	\ifx\next\endList
		\let\next=\relax
	\else
		\ifx\next o
			\put(\columncount,\rowcount){\framebox(1,1){\space}}
			\global\advance\columncount by 1
		\else
			\ifx\next-
				\global\advance\columncount by 1
			\else
				\put(\columncount,\rowcount){\framebox(1,1){$\next$}}
				\global\advance\columncount by 1
			\fi
		\fi
		\let\next=\dorowlist
	\fi
	\next
}

\newcount\ca
\newcount\cb

\def\doprintrowlist{\afterassignment\handlenextprintentry\let\next=}

\def\handlenextprintentry{
	\ifx\next\endList
		\let\next=\relax
	\else
		\put(\ca,\cb){\makebox(1,1){$\scriptstyle\next$}}
		\global\advance\ca by 1
		\let\next=\doprintrowlist
	\fi
	\next
}

\catcode`\@=11

\thinlines
\newskip\Einheit \Einheit=0.6cm
\newcount\xcoord \newcount\ycoord
\newdimen\xdim \newdimen\ydim \newdimen\PfadD@cke \newdimen\Pfadd@cke
\def\PfadDicke#1{\PfadD@cke#1 \divide\PfadD@cke by2 \Pfadd@cke\PfadD@cke \multiply\PfadD@cke by2}
\long\def\LOOP#1\REPEAT{\def\BODY{#1}\ITERATE}
\def\ITERATE{\BODY \let\next\ITERATE \else\let\next\relax\fi \next}
\let\REPEAT=\fi
\def\Punkt{\hbox{\raise-2pt\hbox to0pt{\hss\scriptsize$\bullet$\hss}}}
\def\DuennPunkt(#1,#2){\unskip
  \raise#2 \Einheit\hbox to0pt{\hskip#1 \Einheit
          \raise-2.5pt\hbox to0pt{\hss\normalsize$\bullet$\hss}\hss}}
\def\NormalPunkt(#1,#2){\unskip
  \raise#2 \Einheit\hbox to0pt{\hskip#1 \Einheit
          \raise-3pt\hbox to0pt{\hss\large$\bullet$\hss}\hss}}
\def\DickPunkt(#1,#2){\unskip
  \raise#2 \Einheit\hbox to0pt{\hskip#1 \Einheit
          \raise-4pt\hbox to0pt{\hss\Large$\bullet$\hss}\hss}}
\def\Kreis(#1,#2){\unskip
  \raise#2 \Einheit\hbox to0pt{\hskip#1 \Einheit
          \raise-4pt\hbox to0pt{\hss\Large$\circ$\hss}\hss}}
\def\Diagonale(#1,#2)#3{\unskip\leavevmode
  \xcoord#1\relax \ycoord#2\relax
      \raise\ycoord \Einheit\hbox to0pt{\hskip\xcoord \Einheit
         \unitlength\Einheit
         \line(1,1){#3}\hss}}
\def\AntiDiagonale(#1,#2)#3{\unskip\leavevmode
  \xcoord#1\relax \ycoord#2\relax \advance\xcoord by -0.05\relax
      \raise\ycoord \Einheit\hbox to0pt{\hskip\xcoord \Einheit
         \unitlength\Einheit
         \line(1,-1){#3}\hss}}
\def\Pfad(#1,#2),#3\endPfad{\unskip\leavevmode
  \xcoord#1 \ycoord#2 \thicklines\ZeichnePfad#3\endPfad\thinlines}
\def\ZeichnePfad#1{\ifx#1\endPfad\let\next\relax
  \else\let\next\ZeichnePfad
    \ifnum#1=1
      \raise\ycoord \Einheit\hbox to0pt{\hskip\xcoord \Einheit
         \vrule height\Pfadd@cke width1 \Einheit depth\Pfadd@cke\hss}%
      \advance\xcoord by 1
    \else\ifnum#1=2
      \raise\ycoord \Einheit\hbox to0pt{\hskip\xcoord \Einheit
        \hbox{\hskip-1pt\vrule height1 \Einheit width\PfadD@cke depth0pt}\hss}%
      \advance\ycoord by 1
    \else\ifnum#1=3
      \raise\ycoord \Einheit\hbox to0pt{\hskip\xcoord \Einheit
         \unitlength\Einheit
         \line(1,1){1}\hss}
      \advance\xcoord by 1
      \advance\ycoord by 1
    \else\ifnum#1=4
      \raise\ycoord \Einheit\hbox to0pt{\hskip\xcoord \Einheit
         \unitlength\Einheit
         \line(1,-1){1}\hss}
      \advance\xcoord by 1
      \advance\ycoord by -1
    \fi\fi\fi\fi
  \fi\next}
\def\hSSchritt{\leavevmode\raise-.4pt\hbox to0pt{\hss.\hss}\hskip.2\Einheit
  \raise-.4pt\hbox to0pt{\hss.\hss}\hskip.2\Einheit
  \raise-.4pt\hbox to0pt{\hss.\hss}\hskip.2\Einheit
  \raise-.4pt\hbox to0pt{\hss.\hss}\hskip.2\Einheit
  \raise-.4pt\hbox to0pt{\hss.\hss}\hskip.2\Einheit}
\def\vSSchritt{\vbox{\baselineskip.2\Einheit\lineskiplimit0pt
\hbox{.}\hbox{.}\hbox{.}\hbox{.}\hbox{.}}}
\def\DSSchritt{\leavevmode\raise-.4pt\hbox to0pt{%
  \hbox to0pt{\hss.\hss}\hskip.2\Einheit
  \raise.2\Einheit\hbox to0pt{\hss.\hss}\hskip.2\Einheit
  \raise.4\Einheit\hbox to0pt{\hss.\hss}\hskip.2\Einheit
  \raise.6\Einheit\hbox to0pt{\hss.\hss}\hskip.2\Einheit
  \raise.8\Einheit\hbox to0pt{\hss.\hss}\hss}}
\def\dSSchritt{\leavevmode\raise-.4pt\hbox to0pt{%
  \hbox to0pt{\hss.\hss}\hskip.2\Einheit
  \raise-.2\Einheit\hbox to0pt{\hss.\hss}\hskip.2\Einheit
  \raise-.4\Einheit\hbox to0pt{\hss.\hss}\hskip.2\Einheit
  \raise-.6\Einheit\hbox to0pt{\hss.\hss}\hskip.2\Einheit
  \raise-.8\Einheit\hbox to0pt{\hss.\hss}\hss}}
\def\SPfad(#1,#2),#3\endSPfad{\unskip\leavevmode
  \xcoord#1 \ycoord#2 \ZeichneSPfad#3\endSPfad}
\def\ZeichneSPfad#1{\ifx#1\endSPfad\let\next\relax
  \else\let\next\ZeichneSPfad
    \ifnum#1=1
      \raise\ycoord \Einheit\hbox to0pt{\hskip\xcoord \Einheit
         \hSSchritt\hss}%
      \advance\xcoord by 1
    \else\ifnum#1=2
      \raise\ycoord \Einheit\hbox to0pt{\hskip\xcoord \Einheit
        \hbox{\hskip-2pt \vSSchritt}\hss}%
      \advance\ycoord by 1
    \else\ifnum#1=3
      \raise\ycoord \Einheit\hbox to0pt{\hskip\xcoord \Einheit
         \DSSchritt\hss}
      \advance\xcoord by 1
      \advance\ycoord by 1
    \else\ifnum#1=4
      \raise\ycoord \Einheit\hbox to0pt{\hskip\xcoord \Einheit
         \dSSchritt\hss}
      \advance\xcoord by 1
      \advance\ycoord by -1
    \fi\fi\fi\fi
  \fi\next}
\def\Koordinatenachsen(#1,#2){\unskip
 \hbox to0pt{\hskip-.5pt\vrule height#2 \Einheit width.5pt depth1 \Einheit}%
 \hbox to0pt{\hskip-1 \Einheit \xcoord#1 \advance\xcoord by1
    \vrule height0.25pt width\xcoord \Einheit depth0.25pt\hss}}
\def\Koordinatenachsen(#1,#2)(#3,#4){\unskip
 \hbox to0pt{\hskip-.5pt \ycoord-#4 \advance\ycoord by1
    \vrule height#2 \Einheit width.5pt depth\ycoord \Einheit}%
 \hbox to0pt{\hskip-1 \Einheit \hskip#3\Einheit 
    \xcoord#1 \advance\xcoord by1 \advance\xcoord by-#3 
    \vrule height0.25pt width\xcoord \Einheit depth0.25pt\hss}}
\def\Gitter(#1,#2){\unskip \xcoord0 \ycoord0 \leavevmode
  \LOOP\ifnum\ycoord<#2
    \loop\ifnum\xcoord<#1
      \raise\ycoord \Einheit\hbox to0pt{\hskip\xcoord \Einheit\Punkt\hss}%
      \advance\xcoord by1
    \repeat
    \xcoord0
    \advance\ycoord by1
  \REPEAT}
\def\Gitter(#1,#2)(#3,#4){\unskip \xcoord#3 \ycoord#4 \leavevmode
  \LOOP\ifnum\ycoord<#2
    \loop\ifnum\xcoord<#1
      \raise\ycoord \Einheit\hbox to0pt{\hskip\xcoord \Einheit\Punkt\hss}%
      \advance\xcoord by1
    \repeat
    \xcoord#3
    \advance\ycoord by1
  \REPEAT}
\def\Label#1#2(#3,#4){\unskip \xdim#3 \Einheit \ydim#4 \Einheit
  \def\lo{\advance\xdim by-.5 \Einheit \advance\ydim by.5 \Einheit}%
  \def\llo{\advance\xdim by-.25cm \advance\ydim by.5 \Einheit}%
  \def\loo{\advance\xdim by-.5 \Einheit \advance\ydim by.25cm}%
  \def\o{\advance\ydim by.25cm}%
  \def\ro{\advance\xdim by.5 \Einheit \advance\ydim by.5 \Einheit}%
  \def\rro{\advance\xdim by.25cm \advance\ydim by.5 \Einheit}%
  \def\roo{\advance\xdim by.5 \Einheit \advance\ydim by.25cm}%
  \def\l{\advance\xdim by-.30cm}%
  \def\r{\advance\xdim by.30cm}%
  \def\lu{\advance\xdim by-.5 \Einheit \advance\ydim by-.6 \Einheit}%
  \def\llu{\advance\xdim by-.25cm \advance\ydim by-.6 \Einheit}%
  \def\luu{\advance\xdim by-.5 \Einheit \advance\ydim by-.30cm}%
  \def\u{\advance\ydim by-.30cm}%
  \def\ru{\advance\xdim by.5 \Einheit \advance\ydim by-.6 \Einheit}%
  \def\rru{\advance\xdim by.25cm \advance\ydim by-.6 \Einheit}%
  \def\ruu{\advance\xdim by.5 \Einheit \advance\ydim by-.30cm}%
  #1\raise\ydim\hbox to0pt{\hskip\xdim
     \vbox to0pt{\vss\hbox to0pt{\hss$#2$\hss}\vss}\hss}%
}
\catcode`\@=12

\newtheorem{Theorem}{Theorem}
\newtheorem{Corollary}[Theorem]{Corollary}
\newtheorem{Lemma}[Theorem]{Lemma}

\numberwithin{equation}{section}

\def\P#1{P\big(#1\big)}
\def\Pp#1{P^+\big(#1\big)}
\def\v#1{\left\vert #1\right\vert}
\def\eqref#1{(\ref{#1})}
\def\so{\operatorname{\text {\it so}}}
\def\sp{\operatorname{\text {\it sp}}}

\def\Pf{\operatornamewithlimits{Pf}}
\def\P#1{P\big(#1\big)}
\def\Pp#1{P^+\big(#1\big)}
\def\fl#1{\left\lfloor#1\right\rfloor}

\def\v#1{\left\vert #1\right\vert}
\def\eqref#1{(\ref{#1})}
\def\so{\operatorname{\text {\it so}}}
\def\sp{\operatorname{\text {\it sp}}}

\newtheorem{Proposition}[Theorem]{Proposition}

\begin{document}
\title{ Vicious walkers, friendly walkers and Young tableaux II: With a wall}
\author[C. Krattenthaler]{Christian Krattenthaler{\S}}
\author[A. J. Guttmann]{Anthony J. Guttmann{\dag }}
\author[X. G. Viennot]{\\ Xavier G. Viennot{\ddag }} 
\address{\S Institut f\"{u}r Mathematik, Universit\"{a}t Wien\\
Strudlhofgasse 4, A-1090 Vienna, Austria}
\address{\dag Department of Mathematics and Statistics, The
University of Melbourne\\ Parkville, Victoria 3010, Australia}
\address{\ddag LaBRI, Universit\'{e} Bordeaux 1, 351 cours de la
Lib\'{e}ration, 33405 Talence Cedex, France.
}
%\dedicatory{31st May, 2000}
\begin{abstract}
We derive new results for the number of star
and watermelon configurations of vicious walkers
in the presence of an impenetrable wall by showing 
that these follow from standard results in the
theory of Young tableaux, and combinatorial descriptions
of symmetric functions. For the problem of $n$-friendly
walkers, we derive exact asymptotics for the number of
stars and watermelons both in the absence of a wall and in the
presence of a wall.
\end{abstract}
\maketitle

\section{Introduction}
In an earlier paper \cite{ESG} an expression for the number of
star configurations of vicious walkers on a $d$-dimensional
lattice was obtained,
and the result for the corresponding number of watermelon
configurations was conjectured.
Later, in \cite{GOV} it was shown how certain
results from the theory of Young tableaux, and related
results in algebraic combinatorics enabled one to readily
prove both results.

In this paper, we show how some results from the theory of
symmetric functions can be used to prove analogous results
for the more difficult case of walkers in the presence of
an impenetrable wall. We also give rigorous asymptotic results.

{\em Vicious walkers} 
describes the situation in which two or more walkers arriving
at the same lattice site annihilate one another. Accordingly,
the only configurations we consider are those in which such contacts
are forbidden. Alternatively expressed, we consider mutually 
self-avoiding networks of lattice walks which also model 
directed polymer
networks. The connection of these vicious walker problems to the
5 and  6 vertex model of statistical mechanics was also discussed in \cite{GOV}.

The problem, together with a number of physical applications,
was introduced by Fisher \cite{FIS}. The general model is one
of $p$ random walkers on a 
$d$-dimensional lattice who at 
regular time intervals simultaneously take one step with equal 
probability in the direction of one of the allowed lattice vectors
such that at no time two walkers occupy the same lattice site.

The two standard topologies of interest are that of a \textit{star}
and a \textit{watermelon}. Consider a directed square lattice,
rotated 
$45^\circ$ and augmented by a factor of $\sqrt2$, 
so that the ``unit" vectors on the lattice are
$(1,1)$ and 
$(1,-1)$.
Both configurations consist of $p$ 
branches of length $m$ (the lattice paths along which the walkers proceed) 
which start at $(0,0),(0,2),(0,4),\dots,(0,2p-2).$
The watermelon configurations end at
$(m,k),(m,2+k),(m,4+k),\dots,(m,k+2p-2)$,
for some $k$. For stars, the end-points
of the 
branches all have $x$-coordinate equal to $m,$ but the $y$
 coordinates
are unconstrained, apart from the ordering imposed by the non-crossing
condition. Thus if the end-points are
$(m,e_1),(m,e_2),(m,e_3),\cdots,(m,e_p),$ then 
$e_1 < e_2 < e_3 < \cdots < e_p \le 2p-2+m.$ In the 
problem considered
here, the additional constraint of an impenetrable wall
imposes the condition that at no stage
may any walker step to a point with negative $y$-coordinate.

In \cite{ESG} recurrence relations and the corresponding differential
equations for stars and watermelons on the directed square lattice were
obtained. In the case of watermelons, a 
determinantal form was
evaluated by standard techniques applied to the determinant. In
the case of stars, the results obtained were conjectural, being
equivalent to an earlier conjecture \cite{AME}.
In \cite{GOV} it was shown  how a number of ``standard'' results in the
theory of Young tableaux and partitions lead to a much more intuitive
derivation of the above results, and provided a proof of the
conjectured results.

In recent months a number of authors \cite{Joh00,Pra100,Pra200,Ba00} have
made the fascinating connection between certain properties of two-dimensional
vicious walkers and the eigenvalue distribution of 
certain random
matrix ensembles. In \cite{Joh00} a model is introduced
which can be considered as
a randomly growing Young 
diagram, 
or a totally
asymmetric one-dimensional exclusion process. 
(This could be interpreted in the vicious walker model where at each
time unit {\em exactly one} of the walkers moves. This model occurs already
in \cite{FIS}.)
It is shown that the appropriately scaled shape fluctuations converge
in distribution to the Tracy-Widom distribution of the largest
eigenvalue of the Gaussian Unitary Ensemble (GUE). Similarly, in
\cite{Ba00} a vicious 
walker model is considered in which the end-point
fluctuations of the top-most walker (in our notation) are considered.
In that case the appropriately scaled limiting distribution is
that of the largest eigenvalue of another distribution, the Gaussian
Orthogonal Ensemble (GOE). Finally in \cite{Pra100,Pra200} the height
distribution of a given point in the substrate of a one-dimensional
growth process is considered, and this is generalised to models in
the Kardar-Parisi-Zhang (KPZ) universality class \cite{Ka86}. 
The configurations considered again appear like vicious walkers. Again
fluctuations and other properties of the models are found that follow
GOE or GUE distributions. 
Given that Painlev\'e
transcendents underlie the theory of random matrix ensembles, it would
be of considerable value, and would undoubtedly add greatly
to our understanding of the combinatorial problems we discuss here,
if the connection with random matrix theory and Painlev\'e transcendents 
could be clarified. An important recent development in this clarification
is the the recent paper by Its. {\em et al.} \cite{Itw00}.

%change1
In \cite{GV} and \cite{Katori} two slightly different
generalisations of the vicious walker model
were introduced, both called
the {\em friendly walker model.} In \cite{GV}, the ``vicious''
constraint is systematically relaxed, so that any two walks (but not more than two)
may stay together for
up to $n$ lattice sites in a row, but may never swap sides. We refer
to this as the $n$-friendly walker model. In the limit as $n \rightarrow \infty$
we obtain the $\infty$-friendly walker model  
in which walkers may share an
arbitrary number of steps.
The Tsuchiya-Katori model \cite{Katori}, by contrast, corresponds to a
variant of the $\infty$-friendly walker model which allows any number
of walkers to share any number of bonds, whereas in the Guttmann-V\"oge definition \cite{GV},
only two walkers may share a bond. We subsequently refer to these
two models as the TK and GV models respectively. Thus the number of TK
friendly walk configurations
gives an upper bound on the number of $\infty$-friendly walk conigurations
in the definition of GV. We make use of this
observation in subsequent proofs. Any reference to $n$-friendly walkers
assumes the definition given in \cite{GV}. 
%end of change 
Thus, $n=0$ corresponds to the vicious walker model described
above, and $n = 1$ to the
so-called {\em osculating walker} model in which walkers may touch at a
vertex, but then must part. 
(The osculating walker model is an especially intriguing one, as it
can be seen as a six-vertex model, cf\@. \cite{GOV}, and because, with
special boundary conditions, it produces famous objects in enumerative
combinatorics, alternating sign matrices, cf\@. \cite{BoHaAA, BresAO}.)
%change 2 Line deleted
% end of change 2

Only numerical conjectures for exponents were obtained in \cite{GV} for the
general $n$-friendly walker model. 
Here we provide asymptotics for
this model, which both prove the earlier conjectures, and makes the
earlier results more precise.

In this paper we first re-derive the results of \cite{GOV}  by making
use of Schur functions and {\em odd orthogonal characters}, which
will  
also be needed in our derivation of the results for watermelons
and stars in the presence of a wall.
As well as re-deriving results in the absence of a
wall, we also develop the asymptotics for the number of such configurations.

We then solve the problem in the presence of a wall, or impenetrable barrier
below which the walks may not go. We also give asymptotic results for
both stars and watermelons in this case. We then extend these
results to $m$-friendly stars and watermelons in the presence of a
wall. The problem of vicious walkers in the presence of a wall has
been previously considered by Forrester \cite{Fo89} who set up
the determinantal form, but did not reduce it to a product form.
We accomplish the reduction by the use of {\em symplectic characters}.
A recent preprint \cite{Br00} contains our Theorem~\ref{T3}, 
with an alternative proof.

The makeup of the paper is as follows. In Appendix~A we present
and develop the mathematical tools needed for the asymptotics developed
in the body of the paper. In Appendix~B the 
% change3
two main results on the enumeration of non-intersecting lattice paths
are recalled,
% end of change 3
while Appendix~C is devoted to 
some determinant evaluations. These results will be used throughout
the paper. In 
Section~2 we study stars with fixed end-points, and in 
Section~3 stars with arbitrary end-points, in which we develop many of the
proofs which will be applied {\em mutatis mutandis} in later sections.
Sections~4 and 5 cover the same ground as the two preceding sections,
but this time in the presence of a wall. In Section~6 we discuss watermelon
configurations, and in 
Section~7 we treat the case of watermelons in the
presence of a wall. Section~8 is a short conclusion.

A summary of our results now follows: For the problem of vicious
walkers in the absence of a wall we obtain product forms for the number
of star configurations with fixed end-points, and also for the total number of
stars. For watermelons with given deviation we also derive a product
form, but for the total number of watermelons we are only able to
find the asymptotic form, and not a product form.
By developing the asymptotic form for the total number of 
{\em vicious} stars
(and watermelons), as well as that for the total number of 
{\em $\infty$-friendly}
stars (and watermelons) 
% change4
in the TK model 
% end of change4
we are able to give the asymptotic form for 
$n$-friendly stars and watermelons 
for any $n$.
The number of $n$-friendly stars is 
$\asymp 2^{mp}m^{-p^2/4+p/4}$ as $m$ (the length of the branches)
tends to infinity (see Corollary~\ref{cor:n-friendly-star}). 
It must be expected that it is
asymptotically $c(n)2^{mp}m^{-p^2/4+p/4}$,  
where $c(n)$ is a monotonically increasing
function of $n$. 
Similarly, the number of $n$-friendly watermelons is 
$\asymp 2^{mp}m^{-p^2/2+1/2}$ as $m$ tends to infinity (see
Corollary~\ref{cor:n-friendly-watermelon}), 
and it is expected that 
it is asymptotically 
$f(n)2^{mp}m^{-p^2/2+1/2}$, where $f(n)$ is a monotonically increasing
function of $n$.

Analogous results for systems of walkers
in the presence of a wall are also obtained.
For the problem of vicious
walkers in the presence of a wall we obtain product forms for the number
of star configurations with fixed end-points, and also for the total number of
stars. For watermelons with given deviation we also derive a product
form, but for the total number of watermelons we are only able to
find the asymptotic form, and not a product form.
By developing the asymptotic form for the total number of 
{\em vicious} stars
(and watermelons) in the presence of a wall, as well as that for the 
total number of 
{\em $\infty$-friendly} stars (and watermelons) 
% change 5
in the TK model 
%end of change 5
we are able to give the asymptotic form for 
$n$-friendly stars and watermelons in the presence of a wall
for any $n$.
The number of $n$-friendly stars in the presence of a wall is 
$\asymp 2^{mp}m^{-p^2/2}$ as $m$ tends to infinity (see
Corollary~\ref{cor:n-friendly-star-with}), 
and is expected to be $d(n) 2^{mp}m^{-p^2/2}$, where $d(n)$ 
is a monotonically increasing function of $n.$
Similarly, the number of $n$-friendly watermelons in the presence
of a wall is by $\asymp 2^{mp}m^{-3p^2/4-p/4+1/2}$ as $m$ tends to
infinity (see Corollary~\ref{cor:n-friendly-watermelon-with}), and is
expected to be $g(n)2^{mp}m^{-3p^2/4-p/4+1/2}$,
where $g(n)$ is a monotonically increasing
function of $n.$

\section{Stars with fixed end-points, without a wall.}
A typical star configuration is shown in Figure~\ref{F1}.
Let $e_1<e_2<\dots<e_p$ with $e_i\equiv m\pmod2$, $i=1,2,\dots,p$.
We seek the number of stars with $p$ branches,
the $i$-th branch running from
$A_i=(0,2i-2)$ to $E_i=(m,e_i)$, $i=1,2,\dots,p$. In
Figure~\ref{F1}.a  an example with $p=4$, $m=6$, $e_1=0$, $e_2=2$,
$e_3=6$, $e_4=10$ is given.

\begin{Theorem} \label{T1}
The number of such stars is given by:
\begin{equation} \label{e1}
2^{-\binom p2}\prod _{i=1} ^{p}\frac {(m-i+p)!} {\(\frac {m+e_i} {2}\)!\,\(\frac {m-e_i}
{2}+p-1\)!}\prod _{1\le i<j\le p} ^{}(e_j-e_i).
\end{equation}
\end{Theorem}
\begin{proof}
We describe a proof which uses knowledge about Schur functions. 

Let $\la=(\la_1,\la_2,\dots,\la_m)$ be a partition, i.e., a
nonincreasing sequence of nonnegative integers. Then
the Schur function
$s_\la(x_1,x_2,\dots,x_m)$ is defined by (see \cite[I, (3.1)]{MacdAC}
or \cite[p.~403, (A.4)]{FuHaAA})
\begin{equation} \label{e4}
s_\lambda (x_1, x_2,\dots ,x_m)=\frac {
 \det\limits _{1\le i,j\le m}(x_j^{\lambda _i+m-i})} {
\det\limits_{1\le i,j\le m}(x_j^{m-i})}.
\end{equation}

It is well-known that a combinatorial description of Schur
functions may be given in terms of ({\em semistandard}) {\em tableaux}.
A filling
of the cells of the Ferrers diagram of $\lambda$ with elements of
the set $\{1,2,\dots\}$
which is {\em weakly increasing\/} along rows and {\em strictly increasing\/}
along columns
is called a ({\em semistandard}) {\em tableau of shape $\lambda$\/}.
Figure~\ref{F1}.b shows such a semistandard tableau of shape
$(4,3,2)$.

\begin{figure}[t]
$$
        \Gitter(7,13)(0,-1)
        \Koordinatenachsen(7,13)(0,-1)
        \Pfad(0,0),343443\endPfad
        \Pfad(0,2),334434\endPfad
        \Pfad(0,4),333434\endPfad
        \Pfad(0,6),333334\endPfad
        \DickPunkt(0,0)
        \DickPunkt(0,2)
        \DickPunkt(0,4)
        \DickPunkt(0,6)
        \DickPunkt(6,0)
        \DickPunkt(6,2)
        \DickPunkt(6,6)
        \DickPunkt(6,10)
        \Label\r{\mathit 2}(1,1)
        \Label\r{\mathit 4}(3,1)
        \Label\r{\mathit 5}(4,0)
        \Label\r{\mathit 3}(2,4)
        \Label\r{\mathit 4}(3,3)
        \Label\r{\mathit 6}(5,3)
        \Label\r{\mathit 4}(3,7)
        \Label\r{\mathit 6}(5,7)
        \Label\r{\mathit 6}(5,11)
        \Label\l{A_1}(0,0)
        \Label\l{A_2}(0,2)
        \Label\l{A_3}(0,4)
        \Label\l{A_4}(0,6)
        \Label\r{E_1}(6,0)
        \Label\r{E_2}(6,2)
        \Label\r{E_3}(6,6)
        \Label\r{E_4}(6,10)
\begin{picture}(15,5)(0,0)
\put(12,4){
        \begintableau{3}{4}
        \row{2346}
        \row{446}
        \row{56}
        \endtableau
}
\put(13,2){\small b. A tableau}
\put(3,-3){\small a. A star}
\end{picture}
$$
\vskip.3cm
\caption{}
\label{F1}
\end{figure}

The {\em weight} $\mathbf x^T$
of a tableau $T$ is defined as
\begin{equation}\label{e5}
 \mathbf{x}^{T}:=\prod
x_{T_{i,j}} ,
\end{equation}
where the product is over all entries $T_{ij}$ of $T$.
Given this terminology, the Schur function $s_\la(x_1,x_2,\dots,x_m)$
is also given by (see
\cite[I, (5.12) with $\mu=\emptyset$]{MacdAC}),
\begin{equation}\label{e6}
 s_\lambda(x_1,x_2,\dots,x_m) = \sum_{T}\mathbf{x}^{T},
\end{equation}
where the sum is over all
tableaux $T$ of shape $\lambda$ with entries $\le m$.

In \cite{GOV} it was proved that the number of stars with $p$
branches, as described above, 
can be determined by using
a standard bijection between stars and tableaux, see
Figure~\ref{F1}. First label down-steps by the $x$-coordinate of their
end point, so that a step from $(a-1,b)$ to $(a,b-1)$ is labelled by
$a$, see Figure~\ref{F1}.a. Then, out of the labels of the $j$-th
branch, form the $j$-th column of the corresponding tableau. The
resulting array of numbers is indeed a tableau. This can
be readily seen, since the entries are
trivially strictly increasing along columns, and they are weakly
increasing along rows because the branches do not touch each other.

Thus, given a star with $p$ branches, the $i$-th branch running from
$A_i=(0,2i-2)$ to $E_i=(m,e_i)$, $i=1,2,\dots,p$, one obtains a tableau
with column lengths $\frac {m-e_1}2,\frac {m-e_2}2\dots,\break
\frac {m-e_p}2$.
The {\em shape} (the vector
of row lengths) can be easily extracted from the column lengths.
This correspondence between stars and tableaux is a bijection between
stars with $p$ branches, the $i$-th branch running from
$A_i=(0,2i-2)$ to $E_i=(m,e_i)$, $i=1,2,\dots,p$, and tableaux with
entries at most $m$ and column lengths
$\frac {m-e_1}2,\frac {m-e_2}2\dots,\break\frac {m-e_p}2$.

Clearly, the number of these tableaux is given by \eqref{e6} with
$x_1=x_2=\dots=x_m=1$ and $\la$ the partition whose Ferrers diagram
has column lengths
$\frac {m-e_1}2,\frac {m-e_2}2\dots,\frac {m-e_p}2$.
On the other hand, it is well-known that
(see \cite[I, Sec.~3,  Ex.~1 and 4]{MacdAC},
 \cite[Ex.~A.30, (ii)]{FuHaAA})
\begin{equation} \label{e7}
s_\la(\underbrace{1,1,\dots,1}_m)=\prod _{1\le i<j\le m} ^{}\frac {\la_i-i-\la_j+j}
{j-i}=
\prod _{\rh\in\la} ^{}\frac {m+c_\rh} {h_\rh},
\end{equation}
where $c_\rh$ and $h_\rh$ are the {\em content\/} and the {\em hook
length\/} of the cell $\rh$. The content $c_\rh$ of a cell $\rh=(i,j)$
is $j-i$, whereas the hook-length $h_\rh$ of a cell $\rh$ is the
number of cells in the same row to the right of $\rh$ plus the number
of cells in the same column below $\rh$ plus 1.
The expression obtained can, with some work, be converted
into \eqref{e1}.
\end{proof}

\section{Enumeration of stars with arbitrary end points, without
wall restriction}

The following result was proved in \cite{GOV} using the Bender-Knuth
formula, but no proof of the Bender-Knuth formula was given. As
we require some of the concepts of the proof in subsequent
sections, we briefly repeat the proof given in \cite{GOV}
but 
also review a proof of the Bender-Knuth conjecture.

\begin{Theorem} \label{T2}
The number of stars of length $m$ with $p$ branches equals
\begin{equation} \label{e10}
\prod _{1\le i\le j\le m} ^{}\frac {p+i+j-1} {i+j-1}.
\end{equation}
\end{Theorem}
\begin{proof}Using the correspondence between stars and tableaux
described in the  proof of Theorem~\ref{T1}, we see that we
must count tableaux with entries at most $m$ having at most $p$
columns. This enumeration problem (actually the corresponding
``$q$-enumeration" problem) is known under the name {\em
Bender--Knuth conjecture\/}, and was first proved by B.~Gordon around
1970 (but appeared only much later as \cite{GordAC}). Since then,
many further proofs have been given. See \cite{AndrAJ},
\cite[Th\'eor\'eme~1.1, first identity]{DesaAB},
\cite[I, Sec.~5, Ex.~19]{MacdAC},
\cite[Prop.~7.2]{ProcAD}, \cite{ProcAB},
\cite[Sec.~7]{StemAE} for a selection.
What all these proofs share more or less
explicitly is the following identity, which relates Schur functions
and {\em odd orthogonal characters\/} of the symmetric group 
of rectangular shape,
\begin{equation} \label{e11}
\sum _{\mu,\,\mu_1\le p} ^{}s_\mu(x_1,x_2,\dots,x_m)=(x_1x_2\cdots
x_m)^{p/2}\so_{\big((p/2)^m\big)}
(x_1^{\pm1},x_2^{\pm1},\dots,x_m^{\pm1},1).
\end{equation}
The odd orthogonal characters $\so_\la
(x_1^{\pm1},x_2^{\pm1},\dots,x_m^{\pm1},1)$, where $x_1^{\pm1}$ is 
a shorthand notation
for $x_1,x_1^{-1}$, etc., and 
where $\la$ is an
$m$-tuple $(\la_1,\la_2,\dots,\la_m)$ of integers, or of
half-integers, is defined by
\begin{equation} \label{e12}
\so_\la(x_1^{\pm1},x_2^{\pm1},\dots,x_m^{\pm1},1)=\frac {\det\limits_{1\le i,j\le
m}(x_j^{\la_i+m-i+1/2}-x_j^{-(\la_i+m-i+1/2)})}
{\det\limits_{1\le i,j\le
m}(x_j^{m-i+1/2}-x_j^{-(m-i+1/2)})}.
\end{equation}
Recall that the Schur functions $s_\la(x_1,x_2,\dots,x_m)$ are defined by
\eqref{e4}. While Schur functions are polynomials in
$x_1,x_2,\dots,x_m$ (cf\@. \eqref{e6}), odd orthogonal characters\break
$\so_\la(x_1^{\pm1},x_2^{\pm1},\dots,x_m^{\pm1},1)$ are polynomials
in $x_1,x_1^{-1},x_2,x_2^{-1},\dots,x_m,x_m^{-1},1$. They have a
combinatorial descriptions in terms of certain tableaux as well, see
\cite[Sec.~2]{FuKrAA}, \cite[Sec.~6--8]{ProcAK}, \cite[Theorem~2.3]{SunaAE}.

A variety of different proofs of \eqref{e11} has been given. There
are proofs by a combination of combinatorial and manipulatory
arguments (cf\@.
\cite{GordAC}, \cite[Sec.~7, Cor.~7.4.(a)]{StemAE},
\cite[Theorem~2.3.(1)]{OkadAI}), by use of the theory of
Hall--Littlewood functions (cf\@. \cite[I, Sec.~5, Ex.~16]{MacdAC}),
by use of combinatorial descriptions of
orthogonal characters coming from
algebraic geometry, due to
DeConcini, Procesi, 
Lakshmibai, Musili and Seshadri
(cf\@. \cite[Sec.~7]{ProcAD}, \cite[Theorem~3]{ProcAB},
\cite[Proof of (3.8) in the case $p=0$, stated separately as
(3.12)]{KratBC}). Eventually, a
completely elementary proof was found by Bressoud \cite{BresAN}.

However, what we now require is the evaluation of the left-hand side of
\eqref{e11} at $x_1=x_2=\cdots=x_m=1$,
because this yields, in view of \eqref{e6}, exactly the number of
tableaux under consideration here. In order to evaluate the right-hand
side of \eqref{e11} for $x_1=x_2=\cdots=x_m=1$, 
we may use well-known
formulae for the evaluation of odd orthogonal characters at
these values of the $x_i$'s, namely (see \cite[(24.29)]{FuHaAA},
\cite[Theorem~4.5.(2)]{SunaAE}),
\begin{multline} \label{e13}
\so_\la(1,1,\dots,1)=\prod _{1\le i<j\le m} ^{}\frac {\la_i-i-\la_j+j}
{j-i}\prod _{1\le i\le j\le m}\frac {\la_i+\la_j+2m+1-i-j}
{2m+1-i-j}\\
=\prod _{\rh\in\la} ^{}\frac {2m+1+e_\rh} {h_\rh},
\end{multline}
where, again, $h_\rh$ is the hook length of cell $\rh$, and $e_\rh$
is given by
$$e_\rh=e_{(i,j)}=\begin{cases} \la_i+\la_j-i-j&i\le j,\\
i+j-\la_i'-\la_j'-2&i>j.\end{cases}$$
Here, $\la'$ denotes the partition conjugate to $\la$ (see
\cite[p.~2]{MacdAC} for the definition of conjugate partition).

Using this formula for $\la=\big((p/2)^n\big)$ in \eqref{e11} with
$x_1=x_2=\cdots=x_n=1$, finally leads to \eqref{e10}.
\end{proof}

\begin{Theorem} \label{thm:star-without}
The number of stars with $p$ branches of length $m$
is asymptotically
\begin{equation} \label{eq:star-without}
\begin{cases}
\displaystyle
2^{mp+p^2/4}m^{-p^2/4+p/4}\pi^{-p/4}\bigg(\prod _{\ell=1} ^{p/2}(2\ell-2)!\bigg)(1+O(m^{-1}))&
\text {if $p$ is even,}\\
\displaystyle
2^{mp+p^2/4-1/4}m^{-p^2/4+p/4}\pi^{-p/4+1/4}
\bigg(\prod _{\ell=1} ^{(p-1)/2}(2\ell-1)!\bigg)(1+O(m^{-1}))&
\text {if $p$ is odd,}\end{cases}
\end{equation}
as $m$ tends to infinity.
\end{Theorem}

\begin{proof}We know that the number of stars with $p$ branches of length $m$
is given by the product formula
\begin{equation} \label{eq:num-star}
\prod _{1\le i\le j\le m}
\frac {p+i+j-1} {i+j-1}.
\end{equation}
Therefore, proving \eqref{eq:star-without} amounts to appropriately
rewriting \eqref{eq:num-star} and then applying Stirling's formula.

For convenience, let us introduce the notations $H(n):=\prod _{\ell=1}
^{n}(i-1)!=\prod _{\ell=1} ^{n}(n-i)!$
and $H_2(n):=\prod _{\ell=1} ^{\fl{n/2}}(n-2i)!$. Then the product
\eqref{eq:num-star} can be rewritten as follows,
\begin{align} \notag
\prod _{1\le i\le j\le m}&
\frac {p+i+j-1} {i+j-1}=\prod _{i=1} ^{m}\frac {(p+i+m-1)!\,(2i-2)!}
{(p+2i-2)!\,(i+m-1)!}\\
\label{eq:num-star-H}
&=\frac {H(p+2m)\,H_2(p)\,H_2(2m)\,H(m)}
{H(p+m)\,H_2(p+2m)\,H_2(0)\,H(2m)}.
\end{align}
Our aim is to write this as a product whose range depends only on
$p$. To do so, we need to distinguish between the cases of $p$ being
even or odd.

If $p$ is even, then \eqref{eq:num-star-H} can be written as
$$\prod _{\ell=1} ^{p/2}\frac {(2\ell-2)!} {(2m+2\ell-2)!}
\prod _{\ell=1} ^{p}\frac {(2m+\ell-1)!} {(m+\ell-1)!}.$$
Application of Stirling's formula, and some simplification, yields
the first line of \eqref{eq:star-without}.

If $p$ is odd, then \eqref{eq:num-star-H} can be written as
$$\frac {(p+2m-1)!!} {(p-1)!!}
\prod _{\ell=1} ^{(p+1)/2}\frac {(2\ell-2)!} {(2m+2\ell-2)!}
\prod _{\ell=1} ^{p}\frac {(2m+\ell-1)!} {(m+\ell-1)!}.$$
Renewed application of Stirling's formula, and some simplification, yields
the second line of \eqref{eq:star-without}.
\end{proof}

Next we consider the $\infty$-friendly model for stars.

\begin{Theorem} \label{thm:star-inf-without}
The number of $\infty$-friendly stars
% change 6
 in the TK model 
with $p$ branches of length $m$ is asymptotically
\begin{equation} \label{eq:star-inf-without}
\begin{cases}\displaystyle
2^{mp+3p^2/4-p/2}m^{-p^2/4+p/4}\pi^{-p/4}
\bigg(\prod _{\ell=1} ^{p/2}(2\ell-2)!
\bigg)(1+O(m^{-1}))&
\text {if $p$ is even,}\\
\displaystyle
2^{mp+3p^2/4-p/2-1/4}m^{-p^2/4+p/4}\pi^{-p/4+1/4}\\
\displaystyle
\kern3cm\times
\bigg(\prod _{\ell=1} ^{(p-1)/2}(2\ell-1)!
\bigg)(1+O(m^{-1}))&
\text {if $p$ is odd,}\end{cases}
\end{equation}
as $m$ tends to infinity.
\end{Theorem}
%end of change 6

\begin{proof}
The situation is more difficult here, as we do not have a nice closed
product formula (such as \eqref{eq:num-star}) for $\infty$-friendly
stars. For simplicity, we treat the case of even $m$
only, the case of odd $m$ being completely analogous. 

Consider an $\infty$-friendly star with $p$ branches
of length $m$. It consists of a family $(P_1,P_2,\dots,P_p)$ of
non-crossing lattice paths, $P_i$ running from $A_i=(0,2(i-1))$ to
some point on the line $x=m$, 
$i=1,2,\dots,p$. Figure~\ref{Fig1}.a displays an
example for $p=4$ and $m=6$.

\begin{figure}[t]
$$
        \Gitter(7,13)(0,-1)
        \Koordinatenachsen(7,13)(0,-1)
\raise-2.25pt\hbox{$\Pfad(0,0),343443\endPfad$}
\raise-.75pt\hbox{$\Pfad(0,2),433443\endPfad$}
\raise.75pt\hbox{$\Pfad(0,4),344434\endPfad$}
\raise2.25pt\hbox{$\Pfad(0,6),443334\endPfad$}
        \DickPunkt(0,0)
        \DickPunkt(0,2)
        \DickPunkt(0,4)
        \DickPunkt(0,6)
        \DickPunkt(6,0)
        \DickPunkt(6,2)
%        \DickPunkt(6,4)
        \DickPunkt(6,6)
        \Label\l{A_1}(0,0)
        \Label\l{A_2}(0,2)
        \Label\l{A_3}(0,4)
        \Label\l{A_4}(0,6)
        \Label\r{E_1}(6,0)
        \Label\r{\kern30pt E_2=E_3}(6,2)
%        \Label\r{E_3}(6,4)
        \Label\r{E_4}(6,6)
\hbox{\hskip7cm}
        \Gitter(7,13)(0,-1)
        \Koordinatenachsen(7,13)(0,-1)
        \Pfad(0,0),343443\endPfad
        \Pfad(0,4),433443\endPfad
        \Pfad(0,8),344434\endPfad
        \Pfad(0,12),443334\endPfad
        \DickPunkt(0,0)
        \DickPunkt(0,4)
        \DickPunkt(0,8)
        \DickPunkt(0,12)
        \DickPunkt(6,0)
        \DickPunkt(6,4)
        \DickPunkt(6,6)
        \DickPunkt(6,12)
        \Label\l{\tilde A_1}(0,0)
        \Label\l{\tilde A_2}(0,4)
        \Label\l{\tilde A_3}(0,8)
        \Label\l{\tilde A_4}(0,12)
        \Label\r{\tilde E_1}(6,0)
        \Label\r{\tilde E_2}(6,4)
        \Label\r{\tilde E_3}(6,6)
        \Label\r{\tilde E_4}(6,12)
\hskip6cm$$
\centerline{\small a. An $\infty$-friendly star
\hskip1cm
b. A corresponding family of non-intersecting lattice paths}
\vskip.3cm
\caption{}
\label{Fig1}
\end{figure}
Shifting the $i$-th path, $P_i$, by $2(i-1)$ units up,
we obtain a family $(\tilde P_1,\tilde P_2,\dots,\tilde P_p)$ of
{\em non-intersecting} paths, $\tilde P_i$ running from $(0,4(i-1))$ 
to some point on
the line $x=m$, $i=1,2,\dots,p$, see Figure~\ref{Fig1}.b. 
Clearly, this correspondence is a
bijection. 

% change 7-which runs for 6 pages
The standard way
to find the number of these families of non-intersecting lattice paths
is to resort to Proposition~\ref{prop:Stem} and thus obtain a
Pfaffian for this number. 
However, it seems difficult to derive asymptotic estimates from this
Pfaffian, in particular since Gordon's reductions 
(\cite[implicitly in Sec.~4, 5]{GordAB},
see also \cite[proof of Theorem~7.1]{StemAE}) do not seem to apply. 
Therefore we choose a different path.

For fixed
$e_1,e_2,\dots,e_p$,
the number of families $(\tilde P_1,\tilde P_2,\dots,\tilde P_p)$
of non-intersecting lattice paths, $\tilde P_i$ running from $(0,4(i-1))$
to $(m,2e_i)$
is given by the corresponding Lind\-str\"om--Gessel-Viennot determinant
(see Proposition~\ref{prop:GV}),
\begin{equation} \label{eq:GV-det1A}
\det\limits_{1\le i,j\le p}\(\binom m{\frac {m} {2}+2j-e_i-2}\).
\end{equation}

We have to sum \eqref{eq:GV-det1A} over all $-m/2\le
e_1<e_2<\dots<e_p\le m/2$, and approximate the sum as $m$ tends to
infinity. (It is here where we use the assumption that $m$ is even.
For, any path from a point $(0,4(i-1))$ reaches the vertical line
$x=m$ in a point with even $y$-coordinate.)
We content ourselves to give a rough outline,
as our approach is very much in the spirit of Regev's
asymptotic computation \cite{RegeAF} for Young diagrams in a strip,
and as the proof of Theorem~\ref{thm:watermelon-without} contains a
detailed computation of the same kind, showing all the essentials in
the simpler case of the estimation of a one-fold sum (as
opposed to a $p$-fold sum that we are considering here). As in Regev's
computation, the expression to be estimated is transformed until an
integral is obtained, which then can be evaluated by a limit case of
Selberg's famous integral \cite{SelbAA}. 

To begin with, we bring the determinant \eqref{eq:GV-det1A} into a more
convenient form, by taking out some common factors,
\begin{multline} \label{eq:regdetA}
\det\limits_{1\le i,j\le p}\(\binom m{\frac {m} {2}+2j-e_i-2}\)\\
=\prod _{\ell=1} ^{p}\frac {m!} {(\frac {m} {2}+e_i)!\,(\frac {m}
{2}-e_i+2p-2)!}
\det\limits_{1\le i,j\le p}\Big( (\tfrac {m}
{2}+e_i-2j+3)_{2j-2}\, (\tfrac {m} {2}-e_i+2j-1)_{2p-2j}\Big),
\end{multline}
where $(a)_k$ denotes the standard shifted factorial,
$(a)_k:=a(a+1)\cdots(a+k-1)$, $k\ge1$, $(a)_0:=1$.

The determinant is a polynomial in $m$ and the $e_i$'s. 
It suffices to extract the
leading term, because the contributions of the lower terms to the
overall asymptotics are negligible. In order to do so, 
we observe that, more precisely, the determinant in \eqref{eq:regdetA}
is a polynomial in $m$ and the $e_i$'s of degree $2p^2-2p$ which is
divisible by $\prod _{1\le i<j\le p} ^{}(e_j-e_i)$.

The leading term is
\begin{align*}
&\det\limits_{1\le i,j\le p}\big((\tfrac m2+e_i)^{2j-2}
(\tfrac m2-e_i)^{2p-2j}\big)\\
&\kern10pt=\prod _{i=1} ^{p}(\tfrac m2-e_i)^{2p-2}
\det\limits_{1\le i,j\le p}\(\(\frac {\tfrac m2+e_i} {\tfrac
m2-e_i}\)^{2j-2}\)\\
&\kern10pt=\prod _{i=1} ^{p}(\tfrac m2-e_i)^{2p-2}
\prod _{1\le i<j\le p} ^{}\(\(\frac {\tfrac m2+e_j} {\tfrac
m2-e_j}\)^2
-\(\frac {\tfrac m2+e_i} {\tfrac m2-e_i}\)^2\)\\
&\kern10pt
=\prod _{1\le i<j\le p} ^{}\Big(m(e_j-e_i)\big(\tfrac {m^2}
{2}-2e_ie_j\big)\Big).
\end{align*}
Here we used the Vandermonde determinant evaluation 
to evaluate the determinant in the second line.
On ignoring again terms whose contribution to the overall asymptotics
are negligible, we obtain
\begin{equation} \label{eq:regexprA}
m^{3\binom p2}2^{-\binom p2}\prod _{1\le i<j\le p} ^{}(e_j-e_i)
\end{equation}
as the dominant term in the determinant in \eqref{eq:regdetA}.
Now we have to multiply this expression by the product on the
right-hand side of \eqref{eq:regdetA}, and then
sum the resulting expression over all $-m/2\le e_1<e_2<\dots<e_p\le m/2$. 
In fact, we
may extend the range of summation and sum over all $-m/2\le
e_1\le e_2\le \dots\le e_p\le m/2$, because the expression
\eqref{eq:regexprA} is
zero if any two $e_i$'s should be the same. 

For each $e_i$ separately, the sum over $e_i$ is estimated in the way
as it is done in the proof of Theorem~\ref{thm:watermelon-without} for
the sum over $k$, now using Lemma~\ref{lem:expsum-asy} also for $b$
other than 0. The result is that we obtain
\begin{multline*} 
m^{3\binom p2}2^{-\binom p2}
\prod _{\ell=1} ^{p}\frac {m!} {(\frac {m} {2})!\,(\frac {m} {2}+2p-2)!}
\\
\times\int _{ y_1\le y_2\le \dots\le y_p} ^{}
\bigg(\prod _{1\le i<j\le p} ^{}(y_j-y_i)\bigg)
e^{-\frac {2} {m}\sum _{\ell=1} ^{p}y_i^2}\,dy_1\dots dy_p
\end{multline*}
as an estimation for the number of stars under consideration. In the
integral we perform the substitution $y_i\to x_i\sqrt{m}/2$. This
gives
\begin{multline} \label{eq:SelbA} 
m^{3\binom p2}2^{-\binom p2}
\prod _{\ell=1} ^{p}\bigg(
\frac {m!} {(\frac {m} {2})!\,(\frac {m} {2}+2p-2)!}\bigg)\\
\times\(\frac {\sqrt m} {2}\)^{\binom{p+1}2}
\int _{ x_1\le x_2\le \dots\le x_p} ^{}
\bigg(\prod _{1\le i<j\le p} ^{}\vert x_j-x_i\vert\bigg)
e^{-\frac {1} {2}\sum _{\ell=1} ^{p}x_i^2}\,dx_1\dots dx_p\ .
\end{multline}
At this point, the absolute values in the integrands are
superfluous. However, with the absolute values, the integrand is
invariant under permutations of the $x_i$'s.
Hence, the integral equals
$$\frac {1} {p!}\int _{-\infty} ^{\infty}\cdots \int _{-\infty} ^{\infty}
\bigg(\prod _{1\le i<j\le p} ^{}\vert x_j-x_i\vert\bigg)
e^{-\frac {1} {2}\sum _{\ell=1} ^{p}x_i^2}\,dx_1\dots dx_p\ .$$
This integral is the special case $k=1/2$ 
of Mehta's integral (see \cite[(4.1)]{MacdAF})
\begin{equation*} 
\int _{-\infty} ^{\infty}\cdots \int _{-\infty} ^{\infty}
\bigg(\prod _{1\le i<j\le p} ^{}\vert x_j-x_i\vert^{2k}
\bigg)
e^{-\frac {1} {2}\sum _{\ell=1} ^{p}x_i^2}\,dx_1\dots dx_p
=\(2\pi\)^{p/2}\prod _{\ell=1} ^{p}\frac {(\ell k)!} 
{k!},
\end{equation*}
where $k!$ means $\Gamma(k+1)$ even if $k$ is not an integer.
Substitution of this into \eqref{eq:SelbA} and application of
Stirling's formula to the factorials in the product in
\eqref{eq:SelbA} 
yields \eqref{eq:star-inf-without} after some simplification.
\end{proof}

%end of change 7

Since the expressions \eqref{eq:star-without} and \eqref{eq:star-inf-without} 
are identical except for a
multiplicative constant, we obtain as a consequence
the following result for $n$-friendly models.

\begin{Corollary} \label{cor:n-friendly-star}
As $m$ tends to infinity, $n$-friendly stars with $p$ branches of
length $m$ have, up to a multiplicative
constant, the same asymptotic behaviour,
% change 8
 for the GV as well as for the TK model.
% end of change 8
 More precisely, the number of 
$n$-friendly stars with $p$ branches of length $m$ is $\asymp
2^{mp}m^{-p^2/4+p/4}$, i.e., there are positive constants $c_1$ and $c_2$
such that for large enough $m$ this number is between
$c_12^{mp}m^{-p^2/4+p/4}$ and $c_22^{mp}m^{-p^2/4+p/4}$. 
Under the assumption that there is a constant $c(n)$ such that this
number is asymptotically exactly equal to $c(n)2^{mp}m^{-p^2/4+p/4}$,
then we must have $c(0)<c(1)<c(2)<\cdots$, i.e., for any $n$ there are, 
asymptotically, strictly less $n$-friendly stars with $p$ branches of
length $m$ than $(n+1)$-friendly stars with $p$ branches of
length $m$.\end{Corollary}

\begin{proof}The first assertion follows immediately from
Theorems~\ref{thm:star-without} and
\ref{thm:star-inf-without} since the number of
$n$-friendly stars 
is bounded below by the number of ``genuine"
stars, and is bounded above by the number of $\infty$-friendly
stars 
% change 9
in the TK model.
% end of change 9
So, explicitly, we may choose
$$c_1=\begin{cases}
\displaystyle
2^{p^2/4}\pi^{-p/4}\bigg(\prod _{\ell=1}
^{p/2}(2\ell-2)!\bigg)& 
\text {if $p$ is even,}\\
\displaystyle
2^{p^2/4-1/4}\pi^{-p/4+1/4}
\bigg(\prod _{\ell=1} ^{(p-1)/2}(2\ell-1)!\bigg)&
\text {if $p$ is odd,}\end{cases}$$
and
%change 10
$$c_2=\begin{cases}\displaystyle
2^{3p^2/4-p/2}\pi^{-p/4}
\bigg(\prod _{\ell=1} ^{p/2}(2\ell-2)!\bigg)&
\text {if $p$ is even,}\\
\displaystyle
2^{3p^2/4-p/2-1/4}\pi^{-p/4+1/4}
\bigg(\prod _{\ell=1} ^{(p-1)/2}(2\ell-1)!\bigg)&
\text {if $p$ is odd.}\end{cases}$$

%end of change 10

The second assertion can be proved as follows. Clearly, for any $n$ we have
$c(n)\le c(n+1)$. To see that in fact strict inequality holds, we
identify a set of $(n+1)$-friendly stars which are not $n$-friendly
stars, with the property that its cardinality is
$\asymp (2^{mp}m^{-p^2/4+1/4})$ (as is the cardinality of $n$-friendly stars).
As this set of $(n+1)$-friendly stars we may choose families
$(P_1,P_2,\dots,P_p)$ of paths, such that $P_i$ runs from
$(0,2(i-1))$ through $(2\fl{n/2}+4,2(i-1))$ to the line $x=m$,
$i=1,2,\dots,p$, and $P_1$ and $P_2$
touch each other along $n+1$ consecutive
edges. (This is indeed possible. Let $P_1$ start with an up-step and
$P_2$ start with a down step, then let $P_1$ and $P_2$ go up and down
in parallel for $2\fl{n/2}+2$ steps, then let $P_1$ continue with a
down-step, thus reaching $(2\fl{n/2}+4,0)$,
and $P_2$ continue with an up-step, thus reaching $(2\fl{n/2}+4,2)$.
As $2\fl{n/2}+2\ge n+1$, such paths $P_1$ and
$P_2$ do indeed touch each other along $n+1$ consecutive edges.)

If we disregard the portion of the paths between $x=0$ and
$x=2\fl{n/2}+4$, then what remains is an $(n+1)$-friendly star with
$p$ branches of length $m-2\fl{n/2}-4$. The cardinality of these is
at least the cardinality of ``genuine" stars with
$p$ branches of length $m-2\fl{n/2}-4$, which,
asymptotically, is given by \eqref{eq:star-without} with $m$
replaced by $m-2\fl{n/2}-4$. Up to some constant, this is
$$2^{mp-2\fl{n/2}p-4p}m^{-p^2/4+1/4}(1-(2\fl{n/2}+4)/m)^{-p^2/4+1/4}
(1+O(m^{-1}),
$$
which is $\asymp (2^{mp}m^{-p^2/4+1/4})$,
as desired.
\end{proof}

Clearly, there is abundant evidence that for any fixed $p$ there exist such
constants $c(0),c(1)$, etc. By Theorems~\ref{thm:star-without} and
\ref{thm:star-inf-without} we have computed $c(0)$ and $c(\infty)$. It
appears to be a challenging problem to determine the other constants,
and even just $c(1)$. However, for $p=2$ we can calculate $c(k)$ from the
data given in \cite{GV}, and find $c(k) = \frac{4}{(1 + 2^{-k})\sqrt{\pi}}.$
This result supports our assertion regarding the existence of this increasing
sequence of constants.

\section{Enumeration of stars with fixed end points, with
wall restriction}
For vicious walkers, the previous sections re-derive known results, 
by techniques from algebraic combinatorics. 
We now show how these techniques may be used to
derive new results, for the case of vicious walkers in the
presence of a wall.

\begin{Theorem} \label{T3}
Let $e_1<e_2<\dots<e_p$ with $e_i\equiv m\pmod2$, $i=1,2,\dots,p$.
The number of stars with $p$ branches, the $i$-th branch running from
$A_i=(0,2i-2)$ to $E_i=(m,e_i)$, $i=1,2,\dots,p$, and never going
below the $x$-axis, equals
\begin{equation} \label{e20}
2^{-p^2+p}\prod _{i=1} ^{p}\frac {(e_i+1)\,(m+2i-2)!}
{\(\frac {m+e_i} {2}+p\)!\,\(\frac {m-e_i} {2}+p-1\)!}
\prod _{1\le i<j\le p}
^{}(e_j-e_i)(e_i+e_j+2).
\end{equation}
\end{Theorem}
\begin{proof}
(1) {\it By first principles}. As in \cite{GOV},
we could directly use the main theorem of
non-intersecting lattice paths
(see Proposition~\ref{prop:GV}), to write the
number of stars in question in the form
\begin{equation} \label{e21}
\det_{1\le i,j\le p}\big(\v{\Pp{A_j\to E_i}}\big),
\end{equation}
where $\Pp{A\to E}$ denotes the set of all lattice paths from $A$ to
$E$ that do not go below the $x$-axis. By the ``reflection principle"
(see e.g\@. \cite[p.~22]{ComtAA}), each path number $\v{\Pp{A_j\to E_i}}$
could then be written as a difference of two binomials. It was shown
in \cite[Proof of Theorem~7]{KratAP}, how to evaluate the resulting
determinant (actually, a $q$-analogue was evaluated there). The
evaluation relies on the determinant lemma \cite[Lemma~34]{KratAP}.

However, since we are only interested in plain enumeration there is a
simpler way. As a first step, we may freely attach $2i-2$ up-steps at the
beginning of the $i$-th branch, $i=1,2,\dots,p$, (see Figure~\ref{F2}).
It is obvious that the number of stars with starting points
$A'_i=(-2i+2,0)$ (instead of $A_i=(0,2i-2)$) and end points as in the
statement of the theorem, each branch not going below the $x$-axis (see
Figure~\ref{F2}.b),
is exactly the same as the number of stars
in the statement of the theorem (see Figure~\ref{F2}.a).

\begin{figure}[t]
$$
        \Gitter(8,14)(0,-1)
        \Koordinatenachsen(8,14)(0,-1)
        \Pfad(0,0),3343443\endPfad
        \Pfad(0,2),3334434\endPfad
        \Pfad(0,4),3333434\endPfad
        \Pfad(0,6),3333334\endPfad
        \DickPunkt(0,0)
        \DickPunkt(0,2)
        \DickPunkt(0,4)
        \DickPunkt(0,6)
        \DickPunkt(7,1)
        \DickPunkt(7,3)
        \DickPunkt(7,7)
        \DickPunkt(7,11)
        \Label\l{A_1}(0,0)
        \Label\l{A_2}(0,2)
        \Label\l{A_3}(0,4)
        \Label\l{A_4}(0,6)
        \Label\r{E_1}(7,1)
        \Label\r{E_2}(7,3)
        \Label\r{E_3}(7,7)
        \Label\r{E_4}(7,11)
\hbox{\hskip10cm}
        \Gitter(8,14)(-6,-1)
        \Koordinatenachsen(8,14)(-6,-1)
        \Pfad(0,0),3343443\endPfad
        \Pfad(-2,0),333334434\endPfad
        \Pfad(-4,0),33333333434\endPfad
        \Pfad(-6,0),3333333333334\endPfad
        \DickPunkt(0,0)
        \DickPunkt(-2,0)
        \DickPunkt(-4,0)
        \DickPunkt(-6,0)
        \DickPunkt(7,1)
        \DickPunkt(7,3)
        \DickPunkt(7,7)
        \DickPunkt(7,11)
        \Label\u{A'_1}(0,0)
        \Label\u{A'_2}(-2,0)
        \Label\u{A'_3}(-4,0)
        \Label\u{A'_4}(-6,0)
        \Label\r{E_1}(7,1)
        \Label\r{E_2}(7,3)
        \Label\r{E_3}(7,7)
        \Label\r{E_4}(7,11)
\hskip4cm
$$
\centerline{\small a. A star
\hskip3cm
b. Star with attached ``up-pieces"}
\vskip.3cm
\caption{}
\label{F2}
\end{figure}

If we apply the main theorem of
non-intersecting lattice paths now, then we again obtain a determinant
for the number in question, namely the determinant \eqref{e21} with
$A_j$ replaced by $A_j'$,
\begin{equation} \label{e22}
\det_{1\le i,j\le p}\big(\v{\Pp{A_j'\to E_i}}\big),
\end{equation}
$\Pp{A\to E}$ denoting the set of all lattice paths from $A$ to
$E$ that do not go below the $x$-axis.

Again, by reflection principle, the path number $\v{\Pp{A_j'\to E_i}}$
can be easily computed, so that the determinant \eqref{e22} equals
\begin{equation} \notag
\det_{1\le i,j\le p}\(\frac {e_i+1} {m+2j-1}\binom {m+2j-1}{\frac
{m-e_i} {2}+j-1}\).
\end{equation}
Now we remove as many factors from the determinant as possible. In
that way we obtain
\begin{multline} \label{e23}
(-1)^{\binom p2}
\prod _{i=1} ^{p}\frac {(e_i+1)\,(m+2i-2)!}
{\(\frac {m+e_i} {2}+p\)!\,\(\frac {m-e_i} {2}+p-1\)!}\\
\times
\det_{1\le i,j\le p}\Big(\big(\frac {-m+e_i} {2}-p+1\big)\cdots
\big(\frac {-m+e_i} {2}-j\big)\big(\frac {m+e_i} {2}+p\big)\cdots
\big(\frac {m+e_i} {2}+j-1\big)\Big).\\
\end{multline}
The determinant in \eqref{e23} can clearly be rewritten as
\begin{multline} \notag
\det_{1\le i,j\le p}\Big(\big(\frac {e_i+1} {2}-\frac {m-1} {2}-p\big)\cdots
\big(\frac {e_i+1} {2}-\frac {m-1} {2}-j-1\big)\\
\cdot\big(\frac {e_i+1} {2}
+\frac {m-1} {2}+p\big)\cdots
\big(\frac {e_i+1} {2}-\frac {m-1} {2}+j+1\big)\Big)\\
=\det_{1\le i,j\le p}\(\Big(\Big(\frac {e_i+1} {2}\Big)^2-
\Big(\frac {m-1} {2}-p\Big)^2\Big)\cdots
\Big(\Big(\frac {e_i+1} {2}\Big)^2-\Big(\frac {m-1}
{2}-j-1\Big)^2\Big)\).
\end{multline}
This determinant can be reduced by elementary row manipulations to
$$\det_{1\le i,j\le p}\(\Big(\frac {e_i+1} {2}\Big)^{2(p-j)}\),
$$
which is apparently a Vandermonde determinant and therefore equals
\begin{equation} \notag
\prod _{1\le i<j\le p} ^{}\(\Big(\frac {e_i+1} {2}\Big)^2-\Big(\frac {e_j+1}
{2}\Big)^2\)=2^{-p^2+p}\prod _{1\le i<j\le p}
^{}(e_i-e_j)(e_i+e_j+2).
\end{equation}
Substituting this in \eqref{e23} gives \eqref{e20}.

\medskip
(2) {\it Using knowledge about symplectic characters}. The {\em (even) symplectic
character} $\sp_\la(x_1^{\pm1},x_2^{\pm1},\dots,x_n^{\pm1})$ is
defined by (see \cite[(24.18)]{FuHaAA})
\begin{equation} \label{e24}
\sp_\la(x_1^{\pm1},x_2^{\pm1},\dots,x_n^{\pm1})
=\frac {\det\limits_{1\le i,j\le
n}(x_j^{\la_i+n-i+1}-x_j^{-(\la_i+n-i+1)})}
{\det\limits_{1\le i,j\le
n}(x_j^{n-i+1}-x_j^{-(n-i+1)})}.
\end{equation}
Proctor \cite{ProcAF} also defined {\em odd\/} symplectic characters
$\sp_\la(x_1^{\pm1},x_2^{\pm1},\dots,x_n^{\pm1},1)$,
which are for example defined by
\begin{multline} \label{e24o}
\sp_\la(x_1^{\pm1},x_2^{\pm1},\dots,x_n^{\pm1},1)\\
=\frac {1} {2}\det_{1\le i,j\le n}\big(h_{\la_i-i+j}
(x_1^{\pm1},x_2^{\pm1},\dots,x_n^{\pm1},1)+h_{\la_i-i-j+2}
(x_1^{\pm1},x_2^{\pm1},\dots,x_n^{\pm1},1)\big),
\end{multline}
where $h_k(z_1,z_2,\dots,z_r)=\sum _{1\le i_1\le \dots\le i_m\le r}
^{}z_{i_1}\cdots z_{i_k}$ denotes the {\em $k$-th complete homogeneous
symmetric function\/}.

It is well-known that a combinatorial description of symplectic
characters is given in terms of {\em symplectic tableaux}.
Let $\lambda$ be a partition. A symplectic tableau of shape $\la$ is
a semistandard tableau of shape $\la$ with the additional property
that
\begin{equation} \label{e25}
\text {entries in row $i$ are at least $2i-1$.}
\end{equation}
It is obvious that because of weak increase along rows this condition
may be restricted to the entries in the first column.

Let $n$ be fixed, and let $T$ be a symplectic tableau with entries at
most $2n+1$.
The weight $\mathbf x^T$
of the symplectic tableau $T$ is defined by
\begin{equation}\label{e26}
\mathbf{x}^T = \prod_{l=1}^{n}x_l^{\v{\{T_{i,j} = 2l-1\}} -
        \v{\{T_{i,j} = 2l\}}}.
\end{equation}
where $T_{ij}$ denotes the entry in cell $(i,j)$ of $T$. Note that
entries $2n+1$ do not contribute to the weight.
Given this terminology, the (even) symplectic character
$\sp_\la(x_1^{\pm1},x_2^{\pm1},\dots,x_n^{\pm1})$
is also given by (see
\cite[Theorem~2.3]{SunaAE}, \cite[Theorem~4.2]{ProcAK}),
\begin{equation}\label{e27}
 \sp_\lambda(x_1^{\pm1},x_2^{\pm1},\dots,x_n^{\pm1}) = \sum_{T}\mathbf{x}^{T},
\end{equation}
where the sum is over all symplectic
tableaux $T$ of shape $\lambda$ with entries $\le 2n$, whereas the
odd symplectic character
$\sp_\la(x_1^{\pm1},x_2^{\pm1},\dots,x_n^{\pm1},1)$
is also given by (see
\cite[Theorem~4.2]{ProcAK}),
\begin{equation}\label{e27o}
 \sp_\lambda(x_1^{\pm1},x_2^{\pm1},\dots,x_n^{\pm1},1) = \sum_{T}\mathbf{x}^{T},
\end{equation}
where the sum is over all symplectic
tableaux $T$ of shape $\lambda$ with entries $\le 2n+1$.

The formula for symplectic characters needed here is (see
\cite[(3.27)]{ESKiAA},
\cite[Theorem~4.5.(1)]{SunaAE} and \cite[Prop.~3.2]{ProcAF})
\begin{multline} \label{e28}
\sp_\la(\underbrace{1,1,\dots,1}_m)=\prod _{1\le i<j\le m}
^{}\frac {\la_i-i-\la_j+j}
{j-i}\prod _{1\le i\le j\le m}\frac {\la_i+\la_j+m-i-j+2} {m+2-i-j}\\
=\prod _{\rh\in\la} ^{}\frac {m-f_\rh} {h_\rh},
\end{multline}
where $m$ may be even or odd,
where, again, $h_\rh$ is the hook length of cell $\rh$, and $f_\rh$
is given by
$$f_\rh=f_{(i,j)}=\begin{cases} \la_i'+\la_j'-i-j&i\le j,\\
i+j-\la_i-\la_j-2&i>j.\end{cases}$$
Here again, $\la'$ denotes the partition conjugate to $\la$.

\begin{figure}[t]
$$
        \Gitter(8,14)(0,-1)
        \Koordinatenachsen(8,14)(0,-1)
        \Pfad(0,0),3343443\endPfad
        \Pfad(0,2),3334434\endPfad
        \Pfad(0,4),3333434\endPfad
        \Pfad(0,6),3333334\endPfad
        \DickPunkt(0,0)
        \DickPunkt(0,2)
        \DickPunkt(0,4)
        \DickPunkt(0,6)
        \DickPunkt(7,1)
        \DickPunkt(7,3)
        \DickPunkt(7,7)
        \DickPunkt(7,11)
        \Label\r{\mathit 2}(2,2)
        \Label\r{\mathit 4}(4,2)
        \Label\r{\mathit 5}(5,1)
        \Label\r{\mathit 3}(3,5)
        \Label\r{\mathit 4}(4,4)
        \Label\r{\mathit 6}(6,4)
        \Label\r{\mathit 4}(4,8)
        \Label\r{\mathit 6}(6,8)
        \Label\r{\mathit 6}(6,12)
        \Label\l{A_1}(0,0)
        \Label\l{A_2}(0,2)
        \Label\l{A_3}(0,4)
        \Label\l{A_4}(0,6)
        \Label\r{E_1}(7,1)
        \Label\r{E_2}(7,3)
        \Label\r{E_3}(7,7)
        \Label\r{E_4}(7,11)
\begin{picture}(16,5)(0,0)
\put(15,4){
        \begintableau{3}{4}
        \row{2346}
        \row{446}
        \row{56}
        \put(-1.8,2.3){$1\le$}
        \put(-1.8,1.3){$3\le$}
        \put(-1.8,0.3){$5\le$}
        \endtableau
}
\end{picture}
$$
\centerline{\hskip2cm\small a. A star with a wall
\hskip2cm
b. A symplectic tableau}
\vskip.3cm
\caption{}
\label{F3}
\end{figure}

Now we use a slight variant of the correspondence described in the proof of
Theorem~\ref{T1}. Given a star, we label down-steps by the
$x$-coordinate of their
{\em starting} point, i.e., a step from $(a,b)$ to $(a+1,b-1)$ is labelled by
$a$, see Figure~\ref{F3}.a. Then, again, from the labels of the $j$-th
branch, we form the $j$-th column of the corresponding tableau. It is
evident that the condition that the branches do not go below the
$x$-axis under this correspondence translates exactly into
condition \eqref{e25}. Therefore, in that manner we obtain a
bijection between stars with $p$ branches, the $i$-th branch running from
$A_i=(0,2i-2)$ to $E_i=(m,e_i)$, $i=1,2,\dots,p$, and never going
below the $x$-axis, with symplectic tableaux with entries at most
$m-1$ and with column lengths $\frac {m-e_1}2,\frac {m-e_2}2\dots,\frac
{m-e_p}2$. So, formula
\eqref{e28} solves this enumeration problem and gives \eqref{e20}
upon some manipulation.
\end{proof}

\section{Enumeration of stars with arbitrary end points, with
wall restriction}

\begin{Theorem} \label{T4}
The number of stars of length $m$ with $p$ branches which do not go
below the $x$-axis, and whose end points have $y$-coordinates at
least $s$, $s\equiv m\pmod2$, equals
\begin{equation} \label{e30}
\prod _{i=1} ^{p}\prod _{j=1} ^{(m+s)/2}\prod _{k=1} ^{(m-s)/2}\frac {i+j+k-1} {i+j+k-2}.
\end{equation}
\end{Theorem}

\begin{proof} Using the correspondence between stars 
restricted by a wall and
symplectic tableaux
described in the second proof of Theorem~\ref{T3}, we see that we
want to count symplectic tableaux with entries at most $m-1$ having at most
$(m-s)/2$ rows and at most $p$ columns. This problem was encountered before
by Proctor \cite{ProcAE}. (He was actually interested in enumerating
plane partitions of trapezoidal shape. However, he demonstrates in
\cite{ProcAE} that these are in bijection with the symplectic
tableaux we are considering here.) The solution of the problem lies
in the following identity which relates symplectic characters and
Schur functions of rectangular shape:
\begin{equation} \label{e31}
s_{(c^r)}(\underbrace{1,1,\dots,1}_{N+1})=
\sum _{\nu\subseteq (c^r)} ^{}\sp_\nu(\underbrace{1,1,\dots,1}_N).
\end{equation}
Here $(c^r)$ is short for $(c,c,\dots,c)$, with $r$ occurrences of
$c$. Recall, that in
the argument of a symplectic character
$\sp_\la(x_1^{\pm1},x_2^{\pm1},\dots)$ the term $x_i^{\pm1}$ denotes
the {\em two} arguments $x_i,x_i^{-1}$.

Actually, an identity for ``universal" characters is true, see
\cite[(3.1) summed over all $p$]{KratBC}. This underlying ``universal"
character identity is proved by a combinatorial rule due to
Littlewood \cite{LittAA}, see \cite[Proof of Lemma~4]{ProcAE} and
\cite[Sec.~4]{KratBC}.

Clearly, the formula \eqref{e7}, used in \eqref{e31} with $c=p$,
$r=(m-s)/2$, $N=m-1$, immediately gives
what we want. With some work the resulting expression can be
transformed into \eqref{e30}.
\end{proof}

Formula \eqref{e31} also implies that there should be a bijection
between symplectic tableaux with entries at most $m$ having at most
$s$ rows and at most $p$ columns, on the one hand, and (ordinary)
tableaux with entries at most $m+1$ having $s$ rows and $p$ columns,
on the other hand. Such a bijection has been constructed by Haiman
\cite[Prop.~8.11]{HaimAD}, it is based on Sch\"utzenberger's
\cite{SchuAA} jeu de taquin. Actually, his bijection is between
tableaux of trapezoidal shape (which however are in bijection with
symplectic tableaux as shown by Proctor) and tableaux of rectangular
shape. This bijection could easily be converted into a bijection
between stars of length $m$ with $p$ branches which do not go
below the $x$-axis, and whose end points have $y$-coordinates at
least $m-s$ and watermelons of length $m$ with $p$ branches of deviation
$s$.

\begin{Theorem} \label{thm:star-with}
The number of stars with $p$ branches of length $m$ 
which do not go below the $x$-axis, and whose end points 
have $y$-coordinates at least $s$, $s\equiv m\pmod2$,
is asymptotically
\begin{equation} \label{eq:star-with}2^{mp+p^2-p/2}m^{-p^2/2}\pi^{-p/2}
\bigg(\prod _{\ell=1} ^{p}(\ell-1)!\bigg)(1+O(1/m))
\end{equation}as $m$ tends to infinity.
\end{Theorem}
\begin{proof}
We know that the number of stars with $p$ branches of length $m$
which do not go below the $x$-axis, and whose end points
have $y$-coordinates at least $s$, $s\equiv m\pmod2$,
is given by the product formula\begin{equation} \label{se30}
\prod _{\ell=1} ^{p}
\prod _{i=1} ^{(m+s)/2}\prod _{j=1} ^{(m-s)/2}\frac {i+j+\ell-1} {i+j+\ell-2}=
\prod _{\ell=1} ^{p}\frac {(\ell-1)!\,(m+\ell-1)!} {(\tfrac {m+s}{2}+\ell-1)!\,(\tfrac {m-s} {2}+\ell-1)!}.
\end{equation}
Application of Stirling's formula yields \eqref{eq:star-with} after
a short calculation.
\end{proof}

\begin{Theorem} \label{thm:star-inf-with}
The number of $\infty$-friendly stars
%change 11
 in the TK model
%end of change 11
with $p$ branches of length $m$ 
which do not go below the $x$-axis, and whose end points 
have $y$-coordinates at least $s$, $s\equiv m\pmod2$,
is asymptotically
\begin{equation} \label{eq:star-inf-with}
2^{mp+p^2-3p/2}m^{-p^2/2}\pi^{-p/2}
\bigg(\prod _{\ell=1} ^{p}\frac {(\ell-1)!\,(2\ell-2)!\,(4\ell-2)!} 
{(\ell+p-1)!^2}\bigg)(1+O(m^{-1/2}\log^3m))
\end{equation}
as $m$ tends to infinity.
\end{Theorem}
\begin{proof}
Again, the situation is more difficult here, as we do not have a nice
closed product formula (such as \eqref{eq:star-with}) for
$\infty$-friendly stars. 
%change 12
We shall follow very closely the line of arguments of the proof of
Theorem~\ref{thm:star-inf-without}.
Again, for simplicity, we treat the case of even $m$
and $s=0$ only, other cases being completely analogous. 

As in the proof of Theorem~\ref{thm:star-inf-without},
we begin by transforming 
%end of change 12
$\infty$-friendly
stars into families of non-intersecting lattice paths by shifting the
$i$-th path up by $2(i-1)$ units. Thus,
$\infty$-friendly stars with $p$ branches of length $m$ which do not
go below the $x$-axis are in bijection
with families $(P_1,P_2,\dots,P_p)$ of
non-intersecting lattice paths, $P_i$ running from $(0,4(i-1))$
to $(m,2e_i)$, $i=1,2,\dots,p$, for some integers 
$e_1,e_2,\dots,e_p$ with $0\le e_1<e_2<\dots<e_p$. For fixed
$e_1,e_2,\dots,e_p$,
the number of such families of non-intersecting lattice paths
is given by the corresponding Lind\-str\"om--Gessel-Viennot determinant
(see Proposition~\ref{prop:GV}, and cf\@. also the arguments in the
first proof of Theorem~\ref{T3}, particularly the application
of the reflection principle),
\begin{equation} \label{eq:GV-det1}
\det\limits_{1\le i,j\le p}\(\binom m{\frac {m} {2}+2j-e_i-2}-
\binom m{\frac {m} {2}+2j+e_i-1}\).
\end{equation}

We have to sum \eqref{eq:GV-det1} over all $0\le
e_1<e_2<\dots<e_p\le m/2$, and approximate the sum as $m$ tends to
infinity.
%change13:
As in the proof of Theorem~\ref{thm:star-inf-without},
the expression to be estimated is transformed until an
integral is obtained, which then can be evaluated by a limit case of
Selberg's famous integral \cite{SelbAA}. It is, however, a different
limit case that we need here.
%end of change13:

To begin with, we bring the determinant \eqref{eq:GV-det1} into a more
convenient form, by taking out some common factors,
\begin{multline} \label{eq:regdet}
\det\limits_{1\le i,j\le p}\(\binom m{\frac {m} {2}+2j-e_i-2}-
\binom m{\frac {m} {2}+2j+e_i-1}\)\\
=\prod _{\ell=1} ^{p}\frac {m!} {(\frac {m} {2}+e_i+2p-1)!\,(\frac {m}
{2}-e_i+2p-2)!}\hskip3cm\\
\times \det\limits_{1\le i,j\le p}\Big( (\tfrac {m}
{2}+e_i-2j+3)_{2p+2j-3}\, (\tfrac {m} {2}-e_i+2j-1)_{2p-2j}\\
-(\tfrac {m}
{2}+e_i+2j)_{2p-2j}\, (\tfrac {m} {2}-e_i-2j+2)_{2p+2j-3}\Big),
\end{multline}
where,
%change 14:
as before, 
%end of change 14:
$(a)_k$ denotes the standard shifted factorial,
$(a)_k:=a(a+1)\cdots(a+k-1)$, $k\ge1$, $(a)_0:=1$.

The determinant is a polynomial in $m$ 
%change 15:
and the $e_i$'s. 
%end of change 15:
It suffices to extract the
leading term, because the contributions of the lower terms to the
overall asymptotics are negligible. In order to do so, we consider the
more general determinant
\begin{multline} \label{eq:gen-det3}
\det\limits_{1\le i,j\le p}\big((\tfrac {m} {2}+X_i-2j+\tfrac
{5} {2})_{2p+2j-3}\,
(\tfrac {m} {2}-X_i+2j-\tfrac {1} {2})_{2p-2j}\\
-(\tfrac {m} {2}+X_i+2j-\tfrac {1} {2})_{2p-2j}\,
(\tfrac {m} {2}-X_i-2j+\tfrac {5} {2})_{2p+2j-3}\big).
\end{multline}
Clearly, we regain the determinant in \eqref{eq:regdet} for
$X_i=e_i+1/2$.

The determinant in \eqref{eq:gen-det3} is a polynomial in $m$ and the
$X_i$'s of degree $p(4p-3)$. It is divisible by $\prod _{1\le
i<j\le p} ^{}(X_j^2-X_i^2)\prod _{i=1} ^{p}X_i$.

The leading term of \eqref{eq:gen-det3} is
\begin{align*}
&\det\limits_{1\le i,j\le p}\big((\tfrac m2+X_i)^{2p+2j-3}
(\tfrac m2-X_i)^{2p-2j}-
(\tfrac m2+X_i)^{2p-2j}(\tfrac m2-X_i)^{2p+2j-3}\big)\\
&\kern10pt=\prod _{i=1} ^{p}(\tfrac m2+X_i)^{2p-3/2}(\tfrac
m2-X_i)^{2p-3/2}
\det\limits_{1\le i,j\le p}\(\(\frac {\tfrac m2+X_i} {\tfrac
m2-X_i}\)^{2j-3/2}-
\(\frac {\tfrac m2+X_i} {\tfrac m2-X_i}\)^{-2j+3/2}\)\\
&\kern10pt=\prod _{i=1} ^{p}(\tfrac m2+X_i)^{p-1}(\tfrac m2-X_i)^{3p-2}
\prod _{i=1}
^{p}\(\(\frac {\tfrac m2+X_i} {\tfrac m2-X_i}\)-1\)
\\
&\kern10pt\kern1cm
\times
\prod _{1\le i<j\le p} ^{}\(\(\frac {\tfrac m2+X_i} {\tfrac m2-X_i}\)
-\(\frac {\tfrac m2+X_j} {\tfrac m2-X_j}\)\)
\(1-\(\frac {(\tfrac m2+X_i)} {(\tfrac m2-X_i)}\frac {(\tfrac m2+X_j)}
{(\tfrac m2-X_j)}\)\)\\
&\kern10pt\kern1cm
\times
\so_{(p-1,p-2,\dots,1,0)}
\(\(\tfrac {\tfrac m2+X_1} {\tfrac m2-X_1}\)^{\pm1},\dots,
\(\tfrac {\tfrac m2+X_p} {\tfrac m2-X_p}\)^{\pm1},1\)\\
&\kern10pt=\prod _{1\le i<j\le p} ^{}
\Big(\big((\tfrac m2+X_j)(\tfrac m2-X_i)-
(\tfrac m2+X_i)(\tfrac m2-X_j)\big)\\
&\kern10pt\kern1cm
\cdot
\big((\tfrac m2+X_j)(\tfrac m2+X_i)-(\tfrac m2-X_i)(\tfrac m2-X_j)\big)\Big)
\prod _{i=1} ^{p}\big((\tfrac m2+X_i)-(\tfrac m2-X_i)\big)\\
&\kern.9cm
\times
\prod _{i=1} ^{p}(\tfrac m2+X_i)^{p-1}(\tfrac m2-X_i)^{p-1}
\so_{(p-1,p-2,\dots,1,0)}
\(\(\tfrac {\tfrac m2+X_1} {\tfrac m2-X_1}\)^{\pm1},\dots,
\(\tfrac {\tfrac m2+X_p} {\tfrac m2-X_p}\)^{\pm1},1\).
\end{align*}
Here we used \eqref{eq:unortho} to evaluate the determinant in the
second line.
Now we recall the observation (see the paragraph containing
\eqref{e13}) that the odd orthogonal character is a certain Laurent
polynomial in its variables. In the present context it implies that
the very last line in our computation is a {\em polynomial\/} in the
quantities $(m/2+X_i)$, $(m/2-X_i)$, $i=1,2,\dots,p$, consisting of a
sum of exactly $\so_{(p-1,\dots,1,0)}(1,1,\dots,1)$ monomials, the
evaluation of the orthogonal character at all 1's being given by \eqref{e13}
itself.

Hence, the determinant \eqref{eq:gen-det3} equals
\begin{multline} \label{eq:leading}
\Big(\prod _{1\le i<j\le p} ^{}m(X_j-X_i)m(X_j+X_i)\Big)
\Big(\prod _{i=1} ^{p}2X_i\Big)\\
\times
2^{-p}\bigg(\prod _{\ell=1} ^{p}\frac {(2\ell-2)!\,(4\ell-2)!}
{(\ell+p-1)!^2}\bigg)
\(\(\frac {m} {2}\)^{2p^2-2p}+\cdots\).
\end{multline}

The leading term of the determinant in \eqref{eq:regdet} is
obtained from this
expression under the
substitution of $X_i=e_i+1/2$, $i=1,2,\dots,p$. This substitution
turns \eqref{eq:leading} into
\begin{multline}\label{eq:regexpr}
m^{3p^2-3p}2^{-2p^2+2p}\prod _{1\le i<j\le p} ^{}(e_j-e_i)(e_j+e_i+1)
\prod _{\ell=1} ^{p}(e_i+\tfrac {1} {2})
\prod _{\ell=1} ^{p}\frac
{(2\ell-2)!\,(4\ell-4)!} {(\ell+p-1)!^2}\\
+\text { lower terms}.
\end{multline}
Now we have to multiply this expression by the product on the
right-hand side of \eqref{eq:regdet}, and then
sum the resulting expression over all $0\le e_1<e_2<\dots<e_p\le m/2$. In fact, we
may extend the range of summation and sum over all $0\le
e_1\le e_2\le \dots\le e_p\le m/2$, because the 
%change 16:
expression 
%end of change 16:
\eqref{eq:regexpr} is
zero if any two $e_i$'s should be the same. 

For each $e_i$ separately, the sum over $e_i$ is estimated in the way
as it is done in the proof of Theorem~\ref{thm:watermelon-without} for
the sum over $k$, now using Lemma~\ref{lem:expsum-asy} also for $b$
other than 0. The result is that we obtain
\begin{multline*} 
m^{3p^2-3p}2^{-2p^2+2p}
\prod _{\ell=1} ^{p}\frac {m!} {(\frac {m} {2}+2p-2)!\,(\frac {m} {2}+2p-1)!}
\frac
{(2\ell-2)!\,(4\ell-4)!} {(\ell+p-1)!^2}\\
\times\int _{0\le y_1\le y_2\le \dots\le y_p} ^{}
\bigg(\prod _{1\le i<j\le p} ^{}(y_j-y_i)(y_j+y_i+1)
\prod _{\ell=1} ^{p}(y_i+\tfrac {1} {2})\bigg)
e^{-\frac {2} {m}\sum _{\ell=1} ^{p}y_i^2}\,dy_1\dots dy_p
\end{multline*}
as an estimation for the number of stars under consideration. In the
integral we perform the substitution $y_i\to x_i\sqrt{m}/2$. After
dropping terms which are asymptotically negligible, we obtain
\begin{multline} \label{eq:Selb} 
m^{3p^2-3p}2^{-2p^2+2p}
\prod _{\ell=1} ^{p}\bigg(
\frac {m!} {(\frac {m} {2}+2p-2)!\,(\frac {m} {2}+2p-1)!}
\frac
{(2\ell-2)!\,(4\ell-4)!} {(\ell+p-1)!^2}\bigg)\\
\times\(\frac {m} {4}\)^{\binom{p+1}2}
\int _{0\le x_1\le x_2\le \dots\le x_p} ^{}
\bigg(\prod _{1\le i<j\le p} ^{}\vert x_j^2-x_i^2\vert
\prod _{\ell=1} ^{p}\vert x_i\vert\bigg)
e^{-\frac {1} {2}\sum _{\ell=1} ^{p}x_i^2}\,dx_1\dots dx_p\ .
\end{multline}
At this point, the absolute values in the integrands are
superfluous. However, with the absolute values, the integrand is
invariant under permutations of the $x_i$'s, and under sign changes of
the $x_i$'s. Hence, the integral equals
$$\frac {1} {2^pp!}\int _{-\infty} ^{\infty}\cdots \int _{-\infty} ^{\infty}
\bigg(\prod _{1\le i<j\le p} ^{}\vert x_j^2-x_i^2\vert
\prod _{\ell=1} ^{p}\vert x_i\vert\bigg)
e^{-\frac {1} {2}\sum _{\ell=1} ^{p}x_i^2}\,dx_1\dots dx_p\ .$$
This integral is the special case $k_1+k_3=1$, $k_2=1$ 
of the integral (see \cite[Conj.~6.1, Case~(a)]{MacdAF})
\begin{multline*} 
\int _{-\infty} ^{\infty}\cdots \int _{-\infty} ^{\infty}
\bigg(\prod _{1\le i<j\le p} ^{}\vert x_j^2-x_i^2\vert^{k_2}
\prod _{\ell=1} ^{p}\vert x_i\vert^{k_1+k_3}\bigg)
e^{-\frac {1} {2}\sum _{\ell=1} ^{p}x_i^2}\,dx_1\dots dx_p\\
=\(2^{1-k_1-k_3}\pi\)^{p/2}\prod _{\ell=1} ^{p}\frac {(\frac {1}
{2}\ell k_2)!\,(k_1+k_3+(\ell-1)k_2)!} {(\frac {1}
{2} k_2)!\,(\frac {1} {2}(k_1+k_3+(\ell-1)k_2))!}.
\end{multline*}
Substitution of this into \eqref{eq:Selb} and application of
Stirling's formula to the factorials in the product in \eqref{eq:Selb} 
yields \eqref{eq:star-inf-with} after some simplification.
\end{proof}

\begin{Corollary} \label{cor:n-friendly-star-with}
As $m$ tends to infinity, $n$-friendly stars with $p$ branches of
length $m$ which do not go
below the $x$-axis have, up to a multiplicative
constant, the same asymptotic behaviour,
% change 17
for the GV as
well as for the TK model.
% end of change 17
More precisely, the number of 
$n$-friendly stars with $p$ branches of length $m$ which do not go
below the $x$-axis and whose end points have $y$-coordinates at
least $s$, $s\equiv m\pmod2$, is $\asymp
2^{mp}m^{-p^2/2}$, i.e., there are positive constants $d_1$ and $d_2$
such that for large enough $m$ this number is between
$d_12^{mp}m^{-p^2/2}$ and $d_22^{mp}m^{-p^2/2}$. 
Under the assumption that there is a constant $d(n)$ such that this
number is asymptotically exactly equal to $d(n)2^{mp}m^{-p^2/2}$,
then we must have $d(0)<d(1)<d(2)<\cdots$, i.e., for any $n$ there are, 
asymptotically, strictly less $n$-friendly stars with $p$ branches of
length $m$ than $(n+1)$-friendly stars with $p$ branches of
length $m$.\end{Corollary}

This follows without difficulty from Theorems~\ref{thm:star-with}
and \ref{thm:star-inf-with}, in the same way as
Corollary~\ref{cor:n-friendly-star} follows
from Theorems~\ref{thm:star-without} and
\ref{thm:star-inf-without}.

Clearly, there is abundant evidence that for any fixed $p$ there exist such
constants $d(0),d(1)$, etc. By Theorems~\ref{thm:star-without} and
\ref{thm:star-inf-without} we have computed $d(0)$ and $d(\infty)$. It
appears to be a challenging problem to determine the other constants,
and even just $d(1)$.

\section{Enumeration of watermelons without
wall restriction}

As discussed in the introduction, watermelons are a proper subset of
stars, their end points being 
$(m,k),(m,k+2),(m,k+4),\cdots,(m,k+2p-2).$
The parameter $k$ we refer to as {\em the deviation}.

In \cite{ESG} it was shown that the number of watermelons of length $m$ with
$p$ branches and deviation $k$ 
(where $k\equiv m\pmod 2$) is given by 
(this also follows from
Theorem~\ref{T1} with $e_i=k+2i-2$, $i=1,2,\dots,p$)
\begin{equation} \label{eq:num-watermelon}
\prod _{\ell=1} ^{p}\prod _{i=1} ^{(m+k)/2}\prod _{j=1} ^{(m-k)/2}
\frac {i+j+\ell-1} {i+j+\ell-2}=\prod _{\ell=1} ^{p}
\frac {(m+\ell-1)!\,(\ell-1)!} {(\frac {m+k} {2}+\ell-1)!\, (\frac
{m-k} {2}+\ell-1)!}.
\end{equation}

We 
are not able to give a closed form expression for the total number of
such watermelons
(it is very unlikely that a closed form expression exists in general), 
but we can give an asymptotic result.
\begin{Theorem} \label{thm:watermelon-without}
The number of watermelons with $p$ branches of length $m$ (and
arbitrary deviation)
is asymptotically
\begin{equation} \label{eq:watermelon-without}
2^{mp+p^2-p/2-1/2}m^{-p^2/2+1/2}\pi^{-p/2+1/2}p^{-1/2}
\bigg(\prod _{\ell=1} ^{p}(\ell-1)!\bigg)(1+O(m^{-1/2}\log^3m))
\end{equation}
as $m$ tends to infinity.
\end{Theorem}
\begin{proof}
We know that the number of watermelons with $p$ branches of length $m$
and with deviation $k$ (where $k\equiv m\pmod 2$) is given by
\eqref{eq:num-watermelon}.
Therefore our task is to estimate the sum
\begin{equation} \label{eq:sum-watermelon}
\sum _{k\equiv m \text { (mod } 2)} ^{}\prod _{\ell=1} ^{p}
\frac {(m+\ell-1)!\,(\ell-1)!} {(\frac {m+k} {2}+\ell-1)!\, (\frac
{m-k} {2}+\ell-1)!}.
\end{equation}
We follow the standard way of carrying out such estimations, as
described in \cite[Ex.~5.4]{OdlyAA}.

The dominant terms in the sum on the right-hand side
are those corresponding to $k$'s which
are near $0$. Consequently, we split the sum into three parts,
the terms with ``small" $k$, the terms with ``large" $k$, and the
terms with $k\sim 0$. Let $S(k)$ denote the summand on the
right-hand side of \eqref{eq:sum-watermelon},
$$S(k)=\prod _{\ell=1} ^{p}
\frac {\Gamma(m+\ell)\,\Gamma(\ell)} {\Gamma(\frac {m+k} {2}+\ell)\,
\Gamma(\frac {m-k} {2}+\ell)}.
$$
Then the precise way we split the sum
in \eqref{eq:sum-watermelon} is
\begin{equation} \label{eq:sumsplit}
\sum _{-m\le k\le m} ^{}S(k)=\sum _{-m\le k\le - \sqrt m\log m} S(k)+
\sum _{\sqrt m\log m\le k\le m} S(k)+
\sum _{\v{k}<\sqrt m\log m} ^{}S(k),
\end{equation}
where it is understood without saying that all sums are only over
those $k$ which are of the same parity as $m$.
We will show that the contributions of the first and second term in
\eqref{eq:sumsplit} are negligible, and we will compute the
contribution which the third term provides.

To see that the first term in \eqref{eq:sumsplit} is negligible, it
is enough to observe that every summand $S(k)$ with $-m\le k\le
-\sqrt m\log m$ is bounded above by $S(-\sqrt m\log m)$, and to
compute the asymptotics of $S(-\sqrt m\log m)$ by means of
Stirling's formula,
\begin{align*}
S(-\sqrt m&\log m)=(1+O(1/m))\prod _{\ell=1} ^{p}\Gamma(\ell)
\frac {\(\frac {m} {e}\)^{m}m^{\ell-1}\sqrt{2\pi m}}
{\(\frac {m+\sqrt m\log m} {2e}\)^{\frac {1} {2}(m+\sqrt m\log m)}
\(\frac {m+\sqrt m\log m} {2}\)^{\ell-1}}\\
&\kern1cm
\cdot\frac {1}
{\(\frac {m-\sqrt m\log m} {2e}\)^{\frac {1} {2}(m-\sqrt m\log m)}
\(\frac {m-\sqrt m\log m} {2}\)^{\ell-1}
\sqrt{\frac {1} {2}\pi^2(m^2-m\log^2m)}}\\
&=(1+O(\log^2 m/m))\prod _{\ell=1} ^{p}(\ell-1)!
\frac {2^{m+2\ell-1}} {m^{\ell-1}\sqrt{\pi m}}\\
&\kern1cm
\cdot
\frac {1} {\(1+\frac {\log m} {\sqrt m}\)^{\frac {1} {2}(m+\sqrt m\log
m)+\ell-1}
\(1-\frac {\log m} {\sqrt m}\)^{\frac {1} {2}(m-\sqrt m\log m)+\ell-1}}\\
&=(1+O(\log^2 m/m))\prod _{\ell=1} ^{p}(\ell-1)!
\frac {2^{m+2\ell-1}} {m^{\ell-1}\sqrt{\pi m}}\\
&\kern1cm
\cdot
\frac {1} {\exp\(\(\frac {\log m} {\sqrt m}-\frac {\log^2m} {2m}+O\(\frac
{\log^3 m} {m\sqrt m}\)\)(\frac {1} {2}(m+\sqrt m\log m)+\ell-1)\)}\\
&\kern1cm
\cdot
\frac {1} {\exp\(\(-\frac {\log m} {\sqrt m}-\frac {\log^2m} {2m}+O\(\frac
{\log^3 m} {m\sqrt m}\)\)(\frac {1} {2}(m-\sqrt m\log m)+\ell-1)\)}\\
&=(1+O(\log^3 m/\sqrt m))\prod _{\ell=1} ^{p}(\ell-1)!
\frac {2^{m+2\ell-1}} {m^{\ell-1}\sqrt{\pi m}}e^{-\frac {1} {2}\log^2m}\\
&=(1+O(\log^3 m/\sqrt m))\frac {2^{pm+p^2}}
{m^{p^2/2}\pi^{p/2} }m^{-\frac {p} {2}\log m}
\prod _{\ell=1} ^{p}(\ell-1)!\\
&=O(1/m^3) ,
\end{align*}
because the term $m^{\frac {1} {2}p\log m}$ which appears in the denominator
of the expression in the next-to-last line grows super exponentially.
Therefore, the first term in \eqref{eq:sumsplit} is
$$O\big(\tfrac {1} {2}(m-\sqrt m\log m)/m^3\big)=O(1/m).$$
The second term in \eqref{eq:sumsplit} is equal to the first term,
hence has the same order of magnitude.

\smallskip
Now we turn to the third term in \eqref{eq:sumsplit}. To carry out
our computations, we would have to distinguish between two cases, depending
on whether $m$ is even or odd. The computations in both cases are
however rather similar. Therefore we carry them out in detail just
for the case that $m$ is even and leave it to the reader to complete
the computations for the other case.

In the case that $m$ is even, the third term in
\eqref{eq:sumsplit}, after replacement of $k$ by $2k$, becomes
\begin{equation} \label{eq:term3}
\sum _{\v k\le \frac {1} {2}\sqrt m\log m} ^{}\prod _{\ell=1} ^{p}
\frac {(m+\ell-1)!\,(\ell-1)!} {(m/2+k+\ell-1)!\, (m/2-k+\ell-1)!}.
\end{equation}
For nonnegative $k$, we may rewrite the summand as
\begin{align} \notag
\prod _{\ell=1} ^{p}&
\frac {(m+\ell-1)!\,(\ell-1)!} {(\frac {m} {2}+\ell-1)!^2}
\cdot \frac {(\frac {m} {2}-k+\ell)(\frac {m} {2}-k+\ell+1)\cdot
(\frac {m} {2}+\ell-1)}
{(\frac {m} {2}+\ell)(\frac {m} {2}+\ell+1)\cdot
(\frac {m} {2}+k+\ell-1)}\\
\notag
&=\prod _{\ell=1} ^{p}
\frac {(m+\ell-1)!\,(\ell-1)!} {(\frac {m} {2}+\ell-1)!^2}
\cdot \frac {(1+\frac {-k+\ell} {m/2})(1+\frac {-k+\ell+1} {m/2})\cdot
(1+\frac {\ell-1} {m/2})}
{(1+\frac {\ell} {m/2})(1+\frac {\ell+1} {m/2})\cdot
(1+\frac {k+\ell-1} {m/2})}\\
\label{eq:S(k)-asy}
&=\prod _{\ell=1} ^{p}
\frac {(m+\ell-1)!\,(\ell-1)!} {(\frac {m} {2}+\ell-1)!^2}
\cdot e^{-\frac {2} {m}k^2+O(k^3/m^2)}.
\end{align}
For nonpositive $k$ there is a similar computation which leads to the
same result.

We have to sum the expression \eqref{eq:S(k)-asy}
over $k$ between $-\frac {1} {2}\sqrt m\log m$ and
$\frac {1} {2}\sqrt m\log m$. In that range, the $O(.)$ term is at worst
$O(m^{-1/2}\log^3m)$. Thus, the sum \eqref{eq:term3} turns into
\begin{equation} \label{eq:term3a}
\bigg(\sum _{\v k\le \frac {1} {2}\sqrt m\log m} ^{}
e^{-\frac {2p} {m}k^2}\bigg)
(1+O(m^{-1/2}\log^3m))
\prod _{\ell=1} ^{p}
\frac {(m+\ell-1)!\,(\ell-1)!} {(\frac {m} {2}+\ell-1)!^2}.
\end{equation}
If we extend the sum to run over all integers $k$ then we make an
error which is bounded by $O(1/m)$. The complete sum
$\sum _{k=-\infty} ^{\infty}
e^{-\frac {2p} {m}k^2}$ can be approximated by
Lemma~\ref{lem:expsum-asy} with $b=0$, $N=0$ and $\al=-2p/m$.
The asymptotics of the product
in \eqref{eq:term3a} is easily determined by using Stirling's
formula. Thus we obtain \eqref{eq:watermelon-without}.
\end{proof}

\begin{Theorem} \label{thm:watermelon-inf-without}
The number of $\infty$-friendly watermelons 
% change 18
in the TK model
% end of change 18
with $p$ branches of length $m$ (and arbitrary deviation)
is asymptotically
\begin{equation} \label{eq:watermelon-inf-without}
2^{mp+2p^2-3p/2-1/2}m^{-p^2/2+1/2}\pi^{-p/2+1/2}p^{-1/2}
\bigg(\prod _{\ell=1} ^{p}(\ell-1)!\bigg)(1+O(m^{-1/2}\log^3m))
\end{equation}
as $m$ tends to infinity.
\end{Theorem}
\begin{proof}
The first step is the same as in the proof of
Theorem~\ref{thm:star-inf-without}. We transform $\infty$-friendly
watermelons into families of non-intersecting lattice paths by shifting the
$i$-th path up by $2(i-1)$ units. Thus,
$\infty$-friendly watermelons with $p$ branches of length $m$ and
deviation $k$ are in bijection
with families $(P_1,P_2,\dots,P_p)$ of
non-intersecting lattice paths, $P_i$ running from $(0,4(i-1))$
to $(m,k+4(i-1))$, $i=1,2,\dots,p$.
The number of such families of non-intersecting lattice paths
is given by the corresponding Lind\-str\"om--Gessel-Viennot determinant
(see Proposition~\ref{prop:GV}),
\begin{equation} \label{eq:GV-det}
\det\limits_{1\le i,j\le p}\(\binom m{\frac {m+k} {2}+2i-2j}\).
\end{equation}

Consequently,
our task is to sum \eqref{eq:GV-det} over all $k\equiv m\pmod 2$,
$-m\le k\le m$, and approximate
the sum as $m$ tends to infinity.
The procedure is quite similar to the proof of
Theorem~\ref{thm:watermelon-without}, with the complication that
\eqref{eq:GV-det} cannot be written in closed form.

To begin with, we bring the determinant \eqref{eq:GV-det} into a more
convenient form, by taking out some factors,
\begin{multline} \label{eq:det-factors}
\det\limits_{1\le i,j\le p}\(\binom m{\frac {m+k} {2}+2i-2j}\)=
\prod _{\ell=1} ^{p}\frac {m!} {(\frac {m+k} {2}+2\ell-2)!\,(\frac
{m-k} {2}-2\ell+2p)!}\\
\times\det\limits_{1\le i,j\le p}\big((\tfrac {m+k} {2}+2i-2j+1)_{2j-2}\, (\tfrac
{m-k} {2}-2i+2j+1)_{2p-2j}\big).
\end{multline}

The determinant in \eqref{eq:det-factors} is a polynomial in $m$ and
$k$, of degree at most $4\binom p2$,
\begin{equation} \label{eq:det-expansion}
\det\limits_{1\le i,j\le p}\big((\tfrac {m+k} {2}+2i-2j+1)_{2j-2}\, (\tfrac
{m-k} {2}-2i+2j+1)_{2p-2j}\big)
=\sum _{0\le s+t\le 4\binom p2}
^{}A(s,t)m^sk^t
\end{equation}
say. We claim that, for our asymptotic
considerations, it is sufficient to just consider the leading terms
on the right-hand side of \eqref{eq:det-expansion}. For, consider a
single term $A(s,t)m^sk^t$ on the right-hand side of
\eqref{eq:det-expansion}. It has to be multiplied by the product on
the right-hand side of \eqref{eq:det-factors}, and the resulting
expression is summed over all $k\equiv m\pmod2$, $-m\le k\le m$, so
that one obtains
\begin{equation*}
\sum _{k\equiv m\text { (mod }2)} ^{}A(s,t)m^sk^t
\prod _{\ell=1} ^{p}\frac {m!} {(\frac {m+k} {2}+2\ell-2)!\,(\frac
{m-k} {2}-2\ell+2p)}.
\end{equation*}
This is now handled in the same way as the sum
\eqref{eq:sum-watermelon}. In essence, it is the expression (compare
\eqref{eq:term3a})
\begin{equation*}
\sum _{k} ^{}A(s,t)m^s(2k)^te^{-\frac {2p} {m}k^2}
\prod _{\ell=1} ^{p}\frac {m!} {(\frac {m} {2}+2\ell-2)!\,(\frac
{m} {2}-2\ell+2p)}
\end{equation*}
that needs to be estimated. By \eqref{eq:expsum-asy} with $b=0$, $N=t$ and
$\al=2p/m$, this
is
\begin{multline} \label{eq:single-term}
A(s,t)m^{s+ {t}/ {2}+ {1}/ {2}}
2^{t/2-3/2}p^{-t/2-1/2}\Gamma(\tfrac {t} {2}+\tfrac {1} {2})
(1+O(m^{-1/2}\log^3m))\\
\times
\prod _{\ell=1} ^{p}\frac {m!} {(\frac {m} {2}+2\ell-2)!\,(\frac
{m} {2}-2\ell+2p)}.
\end{multline}
Obviously, the larger $s$ and $t$ are, the larger will be the
contribution of the corresponding term, the largest coming from those
for which $s+t/2$ is maximal.

As it turns out,
the actual degree in $m$ and $k$ of the leading terms of the
determinant in \eqref{eq:det-expansion}
is significantly smaller than $4\binom p2$.
To see what it is, and what the
leading terms are, we consider a more general determinant,
\begin{equation} \label{eq:gen-det1}
\det\limits_{1\le i,j\le p}\big((\tfrac {m} {2}+X_i-2j+1)_{2j-2}\,
(\tfrac {m} {2}-X_i+2j+1)_{2p-2j}\big).
\end{equation}
Clearly, we regain the determinant in \eqref{eq:det-expansion} for
$X_i=2i+k/2$.

The determinant in \eqref{eq:gen-det1} is a polynomial in $m$ and the
$X_i$'s of degree $4\binom p2$. It is divisible by $\prod _{1\le
i<j\le p} ^{}(X_j-X_i)$. Therefore the degree in $m$ and $k$, after
substitution of $X_i=2i+k/2$, $i=1,2,\dots,p$,
of the determinant in \eqref{eq:det-expansion} is at most $3\binom p2$.

The leading term of \eqref{eq:gen-det1} is
\begin{align*}
\det\limits_{1\le i,j\le p}&\big((\tfrac {m} {2}+X_i)^{2j-2}\,
(\tfrac {m} {2}-X_i)^{2p-2j}\big)=(\tfrac {m} {2}-X_i)^{p(2p-2)}
\det\limits_{1\le i,j\le p}\(\(\frac {\tfrac {m} {2}+X_i}
{\tfrac {m} {2}-X_i}\)^{2j-2}\)\\
&=\prod _{1\le i<j\le p} ^{}\((\tfrac {m} {2}+X_j)^2(\tfrac {m}
{2}-X_i)^2-(\tfrac {m} {2}+X_i)^2(\tfrac {m}
{2}-X_j)^2\)\\
&=\prod _{1\le i<j\le p} ^{}(X_j-X_i)\,m\((\tfrac {m} {2}+X_j)(\tfrac {m}
{2}-X_i)+(\tfrac {m} {2}+X_i)(\tfrac {m}
{2}-X_j)\).
\end{align*}
(Clearly, the determinant evaluation used to get from the first to the
second line is the Vandermonde determinant evaluation.)
The leading term of the determinant in \eqref{eq:det-expansion} is this
expression under the
substitution of $X_i=2i+k/2$, $i=1,2,\dots,p$, which equals
$$\prod _{1\le i<j\le p} ^{}(2j-2i)\,m\(\tfrac {m^2} {2}-\tfrac {k^2}
{2}-2ik-2jk-4ij\)=m^{3\binom p2}\prod _{\ell=1}
^{p}(\ell-1)!(1+O(k^2/m^2)).
$$
Hence, in order to compute the asymptotics of the sum of
\eqref{eq:GV-det} over all $k\equiv m\pmod 2$,
it suffices to determine the asymptotics
of the sum over all $k\equiv m\pmod2$ of (recall \eqref{eq:det-factors}
and the remark after \eqref{eq:single-term})
$$
m^{3\binom p2}\prod _{\ell=1} ^{p}\frac {(\ell-1)!\,m!} {(\frac {m+k} {2}+2\ell-2)!\,(
\frac
{m-k} {2}-2\ell+2p)}.
$$
As the considerations leading to \eqref{eq:single-term} showed,
this can be handled completely analogously to the computation
of the asymptotics of the sum \eqref{eq:sum-watermelon}. The result
is exactly \eqref{eq:watermelon-inf-without}.
\end{proof}

Since the expressions \eqref{eq:watermelon-without} and
\eqref{eq:watermelon-inf-without} are identical except for a
multiplicative constant of $2^{p^2-p}$, we obtain as a consequence
the following result for $n$-friendly models.

\begin{Corollary} \label{cor:n-friendly-watermelon}
As $m$ tends to infinity, $n$-friendly watermelons with $p$ branches of
length $m$ (and arbitrary deviation) have, up to a multiplicative
constant, the same asymptotic behaviour,
% change 19
for the GV as
well as for the TK model.
% end of change 19
More precisely, the number of
$n$-friendly watermelons with $p$ branches of length $m$ is $\asymp
2^{mp}m^{-p^2/2+1/2}$, i.e.,
there are positive constants $f_1$ and $f_2$
such that for large enough $m$ this number is between
$f_12^{mp}m^{-p^2/2+1/2}$ and\break $f_22^{mp}m^{-p^2/2+1/2}$.
Under the assumption that there is a constant $f(n)$ such that this
number is asymptotically exactly equal to $f(n)2^{mp}m^{-p^2/2+1/2}$,
then we must have $f(0)<f(1)<f(2)<\cdots$, i.e., for any $n$ there are,
asymptotically, strictly less $n$-friendly watermelons with $p$ branches of
length $m$ than $(n+1)$-friendly watermelons with $p$ branches of
length $m$.
\end{Corollary}

This follows without difficulty from Theorems~\ref{thm:watermelon-without}
and \ref{thm:watermelon-inf-without}, in the same way as
Corollary~\ref{cor:n-friendly-star} follows
from Theorems~\ref{thm:star-without} and
\ref{thm:star-inf-without}.

Clearly, there is abundant evidence that for any fixed $p$ there exist such
constants $f(0),f(1)$, etc. By Theorems~\ref{thm:watermelon-without} and
\ref{thm:watermelon-inf-without} we have computed $f(0)$ and $f(\infty)$. It
appears to be a challenging problem to determine the other constants,
and even just $f(1)$.
However, for $p=2$ we can calculate $f(k)$ from the
data given in \cite{GV}, and find $f(k) = \frac{16}{(1 + 2^{-k})^2\sqrt{\pi}}.$
This result supports our assertion regarding the existence of this increasing
sequence of constants.

\section{Enumeration of watermelons with given deviation, with
wall restriction}

The number of
such watermelons follows immediately from Theorem~\ref{T3} upon setting
$m = n$ and $e_i = k + 2(i-1).$ After a little simplification we arrive at
\begin{Corollary} \label{C1}
The number of watermelons of length $m$ with $p$ branches which do not go
below the $x$-axis, and with deviation $k \ge 0$
 equals
\begin{equation} \label{e60}
\prod _{i=0} ^{\frac{p}{2}-1}\frac{(k+2p-1-2i)!}{(k+2i)!}\prod _{j=0} ^{p-1}
\frac {(m+2j)!j!}{(\frac{m-k}{2}+j)!(\frac{m+k}{2}+j+p)!}.
\end{equation}
An alternative expression follows by elementary manipulation of the above
expression, and is
\begin{equation} \label{eq:num-watermelon-with}
\prod _{\ell=1} ^{p}
\frac {(\ell-1)!\,(k+2\ell-1)_{p-\ell+1}\,(m+2\ell-2)!} {(\frac {m+k}
{2}+\ell+p-1)!\,(\frac {m-k} {2}+\ell-1)!}.
\end{equation}

\end{Corollary}

{}From this second expression we can derive the following theorem
for the asymptotic number of such watermelons, {\em viz.}

\begin{Theorem} \label{thm:watermelon-with}
The number of watermelons with $p$ branches of length $m$ (and
arbitrary deviation) which do not go below the $x$-axis
is asymptotically
\begin{multline} \label{eq:watermelon-with}
2^{mp+9p^2/4-p/4-3/2}m^{-3p^2/4-p/4+1/2}\pi^{-p/2}p^{-p^2/4-p/4-1/2}\\
\times
\Ga\(\tfrac {p^2} {4}+\tfrac {p} {4}+\tfrac {1} {2}\)
\bigg(\prod _{\ell=1} ^{p}(\ell-1)!\bigg)(1+O(m^{-1/2}\log^3m))
\end{multline}
as $m$ tends to infinity.
\end{Theorem}
\begin{proof}
The number of watermelons with $p$ branches of length $m$
and with deviation $k$ (where $k\equiv m\pmod 2$) which do not go
below the $x$-axis is given by \eqref{eq:num-watermelon-with}.
We wish to sum this expression over all $k\ge0$ with $k\equiv
m\pmod 2$ and then approximate it. This is done completely
analogously to the proof of Theorem~\ref{thm:watermelon-without}.
The only difference is that here the sum does not extend to negative
$k$. Again, the dominant terms are those for $k\sim0$. Therefore we
concentrate on the terms for $0\le k\le \sqrt m\log m$. The other
terms are negligible, as is seen in the same way as in the proof of
Theorem~\ref{thm:watermelon-without}. For carrying out the
computations, we would again have to distinguish between the two cases
of $m$ being even or odd. Also here, the computations are rather
parallel. So let us assume in the following that $m$ is even.

If we carry out the computations parallel to those leading from
\eqref{eq:term3} to \eqref{eq:term3a}, then we see that the sum over
all even $k\ge0$ of \eqref{eq:num-watermelon-with} is asymptotically
\begin{multline} \label{eq:term-wat}
\bigg(\sum _{k=0} ^{\infty}
e^{-\frac {2p} {m}k^2}\prod _{\ell=1} ^{p}(2k+2\ell-1)_{p-\ell+1}\bigg)
\prod _{\ell=1} ^{p}
\frac {(\ell-1)!\,(m+2\ell-2)!} {(\frac {m} {2}+\ell+p-1)!\,
(\frac {m} {2}+\ell-1)!}\\
\times
(1+O(m^{-1/2}\log^3m)).
\end{multline}
The sum in this expression can be split into a linear combination of
sums of the form
$\sum _{k=0} ^{\infty}k^Ne^{-\frac {2p} {m} k^2}$. The asymptotics of
the latter are given by Lemma~\ref{lem:expsum-asy}. It implies
particularly that the largest contribution would come from the term
where the exponent $N$ of $k$ is maximal. This term is
$$\sum _{k=0} ^{\infty}2^{\binom {p+1}2}
k^{\binom {p+1}2}e^{-\frac {2p} {m} k^2},$$
which by \eqref{eq:expsum-asy} gives a contribution of $\frac {1}
{2}(m/2p)^{p^2/4+p/4+1/2}\Ga(p^2/4+p/4+1/2)$.
The asymptotics of the product
in \eqref{eq:term-wat} is easily determined by using Stirling's
formula. Putting everything together we obtain \eqref{eq:watermelon-with}.
\end{proof}

Next we consider the $\infty$-friendly model for watermelons with wall
restriction.

\begin{Theorem} \label{thm:watermelon-inf-with}
The number of $\infty$-friendly watermelons
%change 20
in the TK model
%end of change 20
with $p$ branches of length $m$ (and
arbitrary deviation) which do not go below the $x$-axis
is asymptotically
\begin{multline} \label{eq:watermelon-inf-with}
2^{mp+9p^2/4-9p/4-3/2}m^{-3p^2/4-p/4+1/2}\pi^{-p/2}p^{-p^2/4-p/4-1/2}\\
\times
\Ga\(\tfrac {p^2} {4}+\tfrac {p} {4}+\tfrac {1} {2}\)
\bigg(\prod _{\ell=1} ^{p}\frac {(\ell-1)!\,(2\ell-2)!\,(4\ell-2)!}
{(\ell+p-1)!^2}\bigg)(1+O(m^{-1/2}\log^3m))
\end{multline}
as $m$ tends to infinity.
\end{Theorem}

\begin{proof}
As we have already seen, in the context of $\infty$-friendly models
the situation is more difficult, because
we do not have a nice closed
product formula (such as \eqref{eq:num-watermelon-with}) for the
number of $\infty$-friendly watermelons.

The first step is the same as in the proof of
Theorem~\ref{thm:star-inf-without}. We transform $\infty$-friendly
watermelon into families of nonintersecting lattice paths by shifting the
$i$-th path up by $2(i-1)$ units. Thus,
$\infty$-friendly watermelons with $p$ branches of length $m$ and
deviation $k$ which do not go below the $x$-axis are in bijection
with families $(P_1,P_2,\dots,P_p)$ of
nonintersecting lattice paths which do not go below the $x$-axis,
$P_i$ running from $A_i=(0,4(i-1))$ to
$E_i=(m,k+4(i-1))$, $i=1,2,\dots,p$.

The number of these families of nonintersecting lattice paths is
given by the Lind\-str\"om--Gessel--Viennot determinant
(see Proposition~\ref{prop:GV}),
\begin{equation} \label{se21}
\det\limits_{1\le i,j\le p}\big(\v{\Pp{A_j\to E_i}}\big),
\end{equation}
where $\Pp{A\to E}$ denotes the set of all lattice paths from $A$ to
$E$ that do not go below the $x$-axis.
By the ``reflection principle"
(see e.g\@. \cite[p.~22]{ComtAA}), each path number $\v{\Pp{A_j\to E_i}}$
can then be written as a difference of two binomials. Thus we
obtain the determinant
\begin{equation} \label{eq:GV-det-with}
\det\limits_{1\le i,j\le p}\(\binom {m}{\frac {m-k} {2}+2j-2i}-
\binom {m}{\frac {m+k} {2}+2j+2i-3}\)
\end{equation}
for the number of watermelons under consideration.

We have to sum \eqref{eq:GV-det-with} over all $k\equiv m\pmod 2$,
$-m\le k\le m$, and approximate
the sum as $m$ tends to infinity. Having carried out many similar proofs
before, in particular the proof of
Theorem~\ref{thm:watermelon-inf-without}, we have seen
how to approach this problem. As in the proof of
Theorem~\ref{thm:watermelon-inf-without}, we need to determine the
leading terms of \eqref{eq:GV-det-with}.

Once more, we bring the determinant \eqref{eq:GV-det-with} into a more
convenient form, by taking out some factors,
\begin{multline} \label{eq:det-factors3}
\det\limits_{1\le i,j\le p}\(\binom {m}{\frac {m-k} {2}+2j-2i}-
\binom {m}{\frac {m+k} {2}+2j+2i-3}\)\\
\kern-4.5cm
=
\prod _{\ell=1} ^{p}\frac {m!} {(\frac {m+k} {2}+2\ell+2p-3)!\,(\frac
{m-k} {2}-2\ell+2p)!}\\
\times\det\limits_{1\le i,j\le p}\big((\tfrac {m+k} {2}+2i-2j+1)_{2p+2j-3}\,
(\tfrac {m-k} {2}-2i+2j+1)_{2p-2j}\\
-(\tfrac {m+k} {2}+2i+2j-2)_{2p-2j}\,
(\tfrac {m-k} {2}-2i-2j+4)_{2p+2j-3}\big).
\end{multline}
This determinant is exactly the determinant \eqref{eq:gen-det3} with
$X_i=2i+k/2-3/2$. Thus,
the leading term of the determinant in \eqref{eq:det-factors3} is
obtained from \eqref{eq:leading} under the
substitution of $X_i=2i+k/2-3/2$, $i=1,2,\dots,p$. This substitution
turns \eqref{eq:leading} into
\begin{multline*}
m^{3p^2-3p}2^{-p^2+p}\prod _{\ell=1} ^{p}\frac
{(\ell-1)!\,(2\ell-2)!\,(4\ell-4)!} {(\ell+p-1)!^2}
\prod _{\ell=1} ^{p}(2\ell+\tfrac {k} {2}-\tfrac {1} {2})_{p-\ell}
\prod _{\ell=1} ^{p}(2\ell+\tfrac {k} {2}-\tfrac {3} {2})\\
+\text { lower terms}.
\end{multline*}
When expanded, the term in this expression
which will give the largest contribution is
\begin{equation*}
k^{\binom {p+1}2}m^{3p^2-3p}2^{-3p^2/2+p/2}\prod _{\ell=1} ^{p}\frac
{(\ell-1)!\,(2\ell-2)!\,(4\ell-4)!} {(\ell+p-1)!^2}.
\end{equation*}
What remains to do is to multiply this expression by the product on
the right-hand side of \eqref{eq:det-factors3}, then sum the
resulting expression over all $k\ge0$ with $k\equiv m$ mod 2, 
manipulate the summand in a way
analogous to the manipulations in \eqref{eq:S(k)-asy}, and finally
estimate the result using Lemma~\ref{lem:expsum-asy}. After
simplification, the final result is \eqref{eq:watermelon-inf-with}.
\end{proof}

Again, since the expressions \eqref{eq:watermelon-with} and
\eqref{eq:watermelon-inf-with} are identical except for a
multiplicative constant, we obtain as a consequence
the following result for $n$-friendly models.

\begin{Corollary} \label{cor:n-friendly-watermelon-with}
As $m$ tends to infinity, $n$-friendly watermelons with $p$ branches of
length $m$ (and arbitrary deviation) which do not go below the
$x$-axis have, up to a multiplicative
constant, the same asymptotic behaviour, 
%change 21
for the GV as well as for the TK model.
%end of change 21
More precisely, the number of
$n$-friendly watermelons with $p$ branches of length $m$ which do not go
below the $x$-axis is $\asymp
2^{mp}m^{-3p^2/4-p/4+1/2}$, i.e.,
there are positive constants $g_1$ and $g_2$
such that for large enough $m$ this number is between
$g_12^{mp}m^{-3p^2/4-p/4+1/2}$ and $g_22^{mp}m^{-3p^2/4-p/4+1/2}$.
Under the assumption that there is a constant $g(n)$ such that this
number is asymptotically exactly equal to $g(n)2^{mp}m^{-3p^2/4-p/4+1/2}$,
then we must have $g(0)<g(1)<g(2)<\cdots$, i.e., for any $n$ there are,
asymptotically, strictly less $n$-friendly watermelons with $p$ branches of
length $m$ which do not go below the $x$-axis than
$(n+1)$-friendly watermelons with $p$ branches of
length $m$ which do not go below the $x$-axis.
\end{Corollary}

This follows without difficulty from Theorems~\ref{thm:watermelon-with}
and \ref{thm:watermelon-inf-with}, in the same way as
Corollary~\ref{cor:n-friendly-star} follows
from Theorems~\ref{thm:star-without} and
\ref{thm:star-inf-without}. 

Clearly, there is abundant evidence that for any fixed $p$ there exist such
constants $g(0),g(1)$, etc. By Theorems~\ref{thm:watermelon-with} and
\ref{thm:watermelon-inf-with} we have computed $g(0)$ and $g(\infty)$. It
appears to be a challenging problem to determine the other constants,
and even just $g(1)$.

\section{Conclusion}

We have derived new results for the number of star and watermelon configurations
of vicious walkers both in the absence and 
in the presence of an impenetrable wall by showing how these
results follow from standard results in the theory of Young tableaux, and
combinatorial descriptions of symmetric functions. We present the theory of
asymptotic expansions of determinants, and apply this to obtain asymptotic
expressions for the above quantities. We then apply these asymptotic methods
to the broader question of $n$-friendly walkers, both in the presence
and in the absence of a wall, and give asymptotic expansions for stars
and watermelons in all cases. 

\section*{Acknowledgements}
%change 22
We are particularly grateful to John Essam for pointing out problems
in proofs and definitions in an earlier version of the paper.
%end change 22
We would like to thank Richard Brak and Peter Forrester for comments on
the manuscript. One of us (AJG) wishes to acknowledge the financial support
of the Australian Research Council. Another, (XGV) is grateful to the 
Department of Mathematics and Statistics at The University of Melbourne, where
some of this work was carried out. 
Finally, (CK) acknowledges the financial support of the Austrian
Science Foundation FWF, grants P12094-MAT and P13190-MAT.

\bigskip

\appendix

\centerline{\bf Appendix}
\medskip

\global\def\theTheorem{\mbox{A\arabic{Theorem}}}
\setcounter{Theorem}{0}

In this appendix we collect together the tools used in the 
analysis carried out in the body of the paper.
These comprise certain basic asymptotic results, 
the determinantal and Pfaffian formulae 
for the enumeration of nonintersecting lattice paths
with fixed and
with arbitrary endpoints, and
the evaluation of certain determinants and Pfaffians.

\section{Asymptotics}
Aside from Stirling's formula,
the approximation that we use extensively throughout is the following:
\begin{Lemma} \label{lem:expsum-asy}
Let $N$ and $b$ be nonnegative integers. Then, as $\al$ tends to\/ $0^+$,
\begin{equation} \label{eq:expsum-asy}
\sum _{k=b} ^{\infty}k^Ne^{-\al k^2}=
\int _{b} ^{\infty}y^Ne^{-\al y^2}\,dy+O(1)=\frac 
{\Gamma\(\tfrac {N} {2}+\tfrac {1} {2}\)}
{2\,\al^{{N} /{2}+ {1}/ {2}}}
+O(1),
\end{equation}
where the constant in the error term $O(1)$ can be chosen so that it
is independent of $b$.
\end{Lemma}
\begin{proof} 
We first prove \eqref{eq:expsum-asy} for $b=0$.
The Poisson summation theorem (see
\cite[(5.75) with $a=0$]{OdlyAA}, \cite[(2.8.1)]{Titch})
says that
\begin{equation} \label{eq:Poisson}
\sum _{k=-\infty} ^{\infty}f(k)=\sum _{m=-\infty} ^{\infty}
\int _{-\infty} ^{\infty}f(y)e^{-2\pi imy}\,dy,
\end{equation}
for suitable functions $f$. It is for example valid for continuous,
absolutely integrable functions $f$
of bounded variation. The choice of $f(x)=e^{-\al x^2}$ in
\eqref{eq:Poisson} gives \eqref{eq:expsum-asy} for $N=0$ upon little
manipulation. From now on let $N\ge1$.
In \eqref{eq:Poisson} we choose 
\begin{equation} \label{eq:f(x)}
f(x)=\begin{cases} x^N\cdot e^{-\al x^2}&x\ge0,\\
0&x<0.\end{cases}
\end{equation}
This function does satisfy the above mentioned requirements.
Thus we obtain
\begin{align} \notag
\sum _{k=0} ^{\infty}k^Ne^{-\al k^2}
&=\sum _{m=-\infty} ^{\infty}e^{-\pi^2 m^2/\al}
\int _{0} ^{\infty}y^Ne^{-\al(y+\pi im/\al)^2}\,dy\\
\notag
&=\int _{0} ^{\infty}y^Ne^{-\al y^2}\,dy+
\sum _{m=1} ^{\infty}e^{-\pi^2 m^2/\al}
\int _{-\infty} ^{\infty}{\v y}^Ne^{-\al(y+\pi im/\al)^2}\,dy\\
\label{eq:Poisson-appl}
&=\int _{0} ^{\infty}y^Ne^{-\al y^2}\,dy+
\sum _{m=1} ^{\infty}e^{-\pi^2 m^2/\al}
\int _{-\infty} ^{\infty}\v{y-\tfrac {\pi im}\al}^Ne^{-\al y^2}\,dy.
\end{align}
(To justify that we may take the latter integral over real $y$ instead
of over $y$ with imaginary part $\pi im/\al$, it suffices to observe
that the contour integral of the integrand along the rectangle
connecting the extremal points $-M-\pi im/\al$, $M-\pi im/\al$, 
$M+\pi im/\al$, $-M+\pi im/\al$ vanishes, and that the integrals along
the vertical sides of the rectangle tend to zero as $M$ approaches
infinity.)
Next we approximate the integral which appears in the sum. 
We split the integral into two parts,
\begin{equation*} 
\int _{-\infty} ^{\infty}\v{y-\tfrac {\pi im}\al}^Ne^{-\al y^2}\,dy
=\int _{\v y\ge \frac {\pi m}{\al\sqrt 3}}
\v{y-\tfrac {\pi im}\al}^Ne^{-\al y^2}\,dy+
\int _{\v y <\frac {\pi m}{\al\sqrt 3}}
\v{y-\tfrac {\pi im}\al}^Ne^{-\al y^2}\,dy.
\end{equation*}
For the first part, i.e., for $\v y\ge \frac {\pi m}{\al\sqrt 3}$ and
$y$ real, we have
$\v{y-\tfrac {\pi im}\al}\le 2\v y$.
For the second part, i.e., for $\v y< \frac {\pi m}{\al\sqrt 3}$ and
$y$ real, we have
$\v{y-\tfrac {\pi im}\al}\le \tfrac {2\pi m} {\al \sqrt 3}$.
Thus we obtain
\begin{align*} 
\int _{-\infty} ^{\infty}\v{y-\tfrac {\pi im}\al}^Ne^{-\al
y^2}\,dy&\le
\int _{\v y\ge \frac {\pi m}{\al\sqrt 3}}
(2\v{y})^Ne^{-\al y^2}\,dy+
\int _{\v y <\frac {\pi m}{\al\sqrt 3}}
\(\tfrac {2\pi m} {\al\sqrt 3}\)^Ne^{-\al y^2}\,dy\\
&\le 2^N\int _{-\infty} ^{\infty}
\v{y}^Ne^{-\al y^2}\,dy+
\(\tfrac {2\pi m} {\al\sqrt 3}\)^N\int _{-\infty} ^{\infty}
e^{-\al y^2}\,dy.
\end{align*}
The integrals in the last line are easily evaluated by recalling one
of the definitions of the gamma function (see \cite[1.1(1)]{Erdelyi}),
\begin{equation} \label{eq:gamma}
\Gamma(x)=\int _{0} ^{\infty}t^{x-1}e^{-t}\,dt.
\end{equation}
For, substitution of $-\al y^2$ for $t$ and replacement of $x$ by
$(N+1)/2$ yields
\begin{equation*} 
\int _{0} ^{\infty}y^Ne^{-\al y^2}\,dy
=\frac {\Gamma\(\tfrac {N} {2}+\tfrac {1} {2}\)} {2\,\al^{N/2+1/2}}.
\end{equation*}
Combining all this and substituting back into
\eqref{eq:Poisson-appl}, we get
\begin{equation*} 
\sum _{k=0} ^{\infty}k^Ne^{-\al k^2}
=\frac {\Gamma\(\tfrac {N} {2}+\tfrac {1}
{2}\)} {2\,\al^{N/2+1/2}}
+O\(\sum _{m=1} ^{\infty}e^{-\pi^2 m^2/\al}
\(2^{N+1}
\frac {\Gamma\(\tfrac {N} {2}+\tfrac {1} {2}\)} {2\,\al^{N/2+1/2}}+
\(\tfrac {2\pi m} {\al\sqrt 3}\)^N
\frac {\sqrt\pi} {\sqrt\al}\)\).
\end{equation*}
The appearance of the exponential $e^{-\pi^2 m^2/\al}$ makes the $O(.)$
term ``arbitrarily" small. Thus, \eqref{eq:expsum-asy} with $b=0$ follows
immediately.

To establish \eqref{eq:expsum-asy} in full generality, one would
proceed in the same way. One would apply the Poisson summation theorem
\eqref{eq:Poisson} with $g(x)f(x)$ instead of $f(x)$, with $f(x)$ given by
\eqref{eq:f(x)} as before, and $g(x)$
some suitably ``nice" function with $g(x)=0$ for $x\le b-\ep$ and
$g(x)=1$ for $x\ge b$, $\ep\ge0$. Everything else is completely analogous.
The result is then obtained by letting $\ep\to0$.
\end{proof}

\section{The enumeration of nonintersecting lattice paths}
If one wants to enumerate nonintersecting lattice paths with given
starting and end points, then the solution is given by the
Lind\-str\"om--Gessel--Viennot determinant \cite[Lemma~1]{LindAA},
\cite[Corollary~2]{GeViAB} (cf\@. also \cite{KaMGAC} and \cite{KarlAB}
for continuous and probabilistic versions of the same problem),
\begin{Proposition} \label{prop:GV}
Let $A_1,A_2,\dots,A_p$ and  $E_1,E_2,\dots,E_p$ be lattice points, 
with the property that if $1\le i<j\le p$ and $1\le k<l\le p$,
then any path from $A_i$ to $E_l$ must intersect any path from
$A_j$ to $E_k$. Then
the number of families $(P_1,P_2,\dots,P_p)$ of nonintersecting
lattice paths, where $P_i$ runs from $A_i$ to $E_i$, $i=1,2,\dots,p$,
is given by
\begin{equation} \label{eq:GV}
\det_{1\le i,j\le p}\big(\v{\P{A_j\to E_i}}\big),
\end{equation}
where $\P{A\to E}$ denotes the set of all lattice paths from $A$ to
$E$.

\end{Proposition}

If, however, we want to enumerate
nonintersecting lattice paths with given starting points, but where
the end points may be any points from a given {\em set\/} of points,
then the solution can be given in terms of a Pfaffian, as was shown
by Okada \cite[Theorem~3]{OkadAA} and Stembridge \cite[Theorem~3.1]{StemAE}.

\begin{Proposition} \label{prop:Stem}
Let $p$ be even. Let $A_1,A_2,\dots,A_p$ be lattice points, and let
$\mathbf E=\{E_i:i\in I\}$ be a set of lattice points, 
where $I$ is a linearly ordered set of
indices, with the property that if $1\le i<j\le p$ and $k,l\in I$,
$k<l$, then any path from $A_i$ to $E_l$ must intersect any path from
$A_j$ to $E_k$. Then
the number of families $(P_1,P_2,\dots,P_p)$ of nonintersecting
lattice paths, where $P_i$ runs from $A_i$ to some point in the set
$I$, is given by
\begin{equation} \label{eq:Stem}
\Pf\limits_{1\le i<j\le p}(Q_{\mathbf E}(A_i,A_j)),
\end{equation}
where $Q_{\mathbf E}(A_i,A_j)$ is the number of all pairs of
nonintersecting lattice paths, one connecting $A_i$ to $\mathbf E$,
the other connecting $A_j$ to $\mathbf E$.
\end{Proposition}

% change 23
\section{Some determinants}
% end of change 23
In our computations we need the following determinant evaluations.
All of them are readily proved by
the standard argument that proves Vandermonde-type determinant
evaluations.

\begin{Lemma} \label{lem:dets}
Let $N$ by a nonnegative integer. Then
\begin{align} \label{eq:ortho1}
\det\limits_{1\le i,j\le N}\(x_i^{j}+x_i^{1-j}\)&=
(x_1x_2\cdots x_N)^{1-N}
\prod _{1\le i<j\le N} ^{}(x_i-x_j)(1-x_ix_j)\prod _{i=1}
^{N}(x_i+1),\\
\label{eq:sympl}
\det\limits_{1\le i,j\le N}\(x_i^{j}-x_i^{-j}\)&=
(x_1x_2\cdots x_N)^{-N}
\prod _{1\le i<j\le N} ^{}(x_i-x_j)(1-x_ix_j)\prod _{i=1}
^{N}(x_i^2-1)\\
\label{eq:ortho2}
\det\limits_{1\le i,j\le N}\(x_i^{j-1/2}-x_i^{-j+1/2}\)&=
(x_1x_2\cdots x_N)^{-N+1/2}
\prod _{1\le i<j\le N} ^{}(x_i-x_j)(1-x_ix_j)\prod _{i=1}
^{N}(x_i-1).
\end{align}
\end{Lemma}

In the proof of Theorem~\ref{thm:watermelon-inf-with}
we need to express another, very similar determinant in terms of odd
orthogonal characters. The latter are defined by
\eqref{e12}.
A combination of \eqref{e12} with $\la=(N-1,N-2,\dots,1,0)$ 
and \eqref{eq:ortho2} gives
\begin{multline} \label{eq:unortho}
\det\limits_{1\le i,j\le N}\(x_i^{2j-3/2}-x_i^{-2j+3/2}\)=
\det\limits_{1\le i,j\le N}\(x_i^{j-1/2}-x_i^{-j+1/2}\)\\
\kern4cm\times
\so_{(N-1,N-2,\dots,1,0)}(x_1^{\pm1},x_2^{\pm1},\dots,x_N^{\pm1},1)\\
=(x_1x_2\cdots x_N)^{-N+1/2}
\prod _{1\le i<j\le N} ^{}(x_i-x_j)(1-x_ix_j)\prod _{i=1}
^{N}(x_i-1)\\
\times
\so_{(N-1,N-2,\dots,1,0)}(x_1^{\pm1},x_2^{\pm1},\dots,x_N^{\pm1},1).
\end{multline}

%change 24-large slabs eliminated
% end of change 24

\end{document}